\documentclass[12pt]{article}

\usepackage{iip-arxiv-refactored}

\setcitestyle{authoryear,round}

\usepackage{amsmath}
\usepackage{graphicx}
\usepackage{multirow}
\usepackage{mathtools}
\usepackage{physics}
\usepackage{tabularx}
\usepackage{makecell}
\usepackage{subcaption}
\usepackage{amssymb}
\usepackage{amsthm}
\usepackage{braket}

\usepackage{booktabs}
\usepackage{siunitx}
\usepackage{threeparttable}
\usepackage{bm}

\usepackage{xcolor}
\usepackage{tikz}
\usepackage{quantikz}
\usetikzlibrary{decorations.pathmorphing,arrows.meta,patterns,quantikz2}

\newcommand{\envdim}{N_{\mathrm e}}
\newcommand{\bx}{\bm{x}}
\newcommand{\bw}{\bm{w}}
\newcommand{\br}{\bm{r}}
\newcommand{\bsigma}{\bm{\sigma}}
\newcommand{\balpha}{\bm{\alpha}}
\newcommand{\halpha}{\hat\alpha}
\newcommand{\bhalpha}{\bm{\halpha}}
\newcommand{\btheta}{\bm{\theta}}
\newcommand{\IQC}{\texttt{IQC}}
\newcommand{\IQCail}{\texttt{IQC-AIL}}
\newcommand{\IQCalpha}{\texttt{IQC-$\alpha$}}
\newcommand{\IQCmulti}[1][]{%
  \texttt{IQC-Multi\if\relax\detokenize{#1}\relax\else-#1\fi}%
}

\theoremstyle{plain}

\theoremstyle{definition}

\theoremstyle{remark}

\title{Fourier Analysis of Parametrized Interactive Quantum Classifiers}

\author[a]{Fábio Novaes,}
\affiliation[a]{Academic Unit of Belo Jardim, Federal Rural University of Pernambuco, Belo Jardim, Pernambuco 55156-580, Brazil}
\emailAdd{fabio.novaes@ufrpe.br}
\author[b]{Fernando M. de Paula Neto}
\affiliation[b]{Center of Informatics, Federal University of Pernambuco, Recife, Pernambuco 50740-560, Brazil}
\emailAdd{fernando@cin.ufpe.br}
\author[b]{and João V. M. Cardoso}
\emailAdd{jvmc@cin.ufpe.br}

\abstract{ Interactive Quantum Classifiers (\IQC{}s) constitute a
  family of quantum machine learning models inspired by open quantum
  systems, in which the interaction between a target qubit and an
  environment is described by a Hamiltonian. Previous works introduced
  alternative Hamiltonian parameterizations and showed empirically
  that they can improve classification performance, but the role of
  these parameters in the resulting classifier remains poorly
  understood. In this work, we derive a closed-form expression for the
  reduced quantum channel generated by a parametrized \IQC{} with a
  single target qubit. The analytical solution explicitly reveals how
  the Hamiltonian parameters control the constant, sine, and cosine
  components of the classifier output, establishing a Fourier
  interpretation of the induced feature map. This analysis motivates a
  generalized family of Hamiltonian encodings, including
  matrix-parameterized environmental Hamiltonians whose Fourier
  components depend on linear combinations of input features, thereby
  enabling non-separable Fourier structures. Numerical experiments on
  synthetic and real-world datasets show that the proposed models can
  improve classification performance on several nonlinear
  benchmarks. The generalized matrix encoding achieves the strongest
  aggregate performance in the evaluated benchmark, while a simpler
  four-parameter extension often attains comparable performance with
  substantially fewer trainable parameters. We additionally
  characterize the generated state ensembles using the standard
  fidelity-based expressibility measure, finding that global
  expressibility does not directly predict classification
  performance. Our results provide an analytical characterization of
  parametrized Hamiltonians in Interactive Quantum Classifiers and
  establish Fourier analysis as a useful framework for understanding
  and designing open-system-inspired quantum learning models.  }

\keywords{Quantum Machine Learning, Interactive Quantum classifiers, Fourier Analysis, Open Quantum Systems}

\begin{document}
\maketitle
\section{Introduction}\label{sec:introduction}

Parameterized quantum circuits (PQCs) are among the principal
frameworks for quantum machine learning on near-term quantum hardware.
Their widespread use has motivated theoretical studies of the expressive
power, trainability, and approximation capabilities of quantum models
beyond empirical performance alone.

A key step toward this understanding was the Fourier characterization
of PQC-based quantum models. \cite{schuldEffectDataEncoding2021}
showed that expectation-value models can be written as partial Fourier
series in the input features, with accessible frequencies determined
by the eigenvalue differences of the Hamiltonian generators used for
data encoding, while variational circuit parameters and measurement
operators determine the corresponding Fourier coefficients. This
framework established a direct connection between Hamiltonian design
and the function class represented by a quantum model. Subsequent work
has investigated how different circuit architectures learn Fourier
series, including dissipative quantum neural networks
\citep{heimannLearningFourierSeries2025}, extended the framework to
multidimensional Fourier expansions
\citep{casasMultidimensionalFourierSeries2023a}, and introduced
trainable-frequency architectures in which the generator eigenspectrum
can change during optimization \citep{jaderbergLetQuantumNeural2023}.

A complementary line of work has investigated quantum learning
architectures based on ancillary systems, reduced dynamics, and
open-system evolution. Early approaches to quantum classification used
open-system dynamics and environmental degrees of freedom to process
classical information, including steady-state classifiers based on
reservoir-induced dynamics \citep{Tur18} and dissipative classifiers
involving auxiliary quantum systems \citep{Wan23,Kor23}. In parallel,
quantum reservoir computers have been studied in terms of the
nonlinearity of their input--output maps and the expressive
limitations imposed by input encoding
\citep{mujalAnalyticalEvidenceNonlinearity2021,schutteExpressivityQuantumReservoir2026}.
More recently, dissipative quantum neural networks have also been
investigated for their learning and Fourier-representation
capabilities \citep{Sha20,heimannLearningFourierSeries2025}, while
trainable quantum channels have explored engineered dissipation and
non-unitary dynamics as computational resources \citep{Wen26}.

Despite these connections, these perspectives have rarely been
combined explicitly. Fourier analyses of parameterized quantum
circuits have characterized the spectral structure of their
input--output functions, while quantum reservoir computers have been
studied primarily in terms of nonlinear processing, memory, and
expressivity. Open-system quantum learning models, in turn, have
generally been investigated through dissipative mechanisms,
steady-state behavior, trainability, or channel-level
properties. Although trainable-frequency models demonstrate that
trainable encoding-Hamiltonian parameters can control the Fourier
spectrum of unitary quantum models
\cite{jaderbergLetQuantumNeural2023}, and trainable-channel models
demonstrate spectral modulation at the level of effective observables
\cite{Wen26}, these works do not provide an explicit characterization
of how the eigenvalues of an input-dependent environmental
Hamiltonian, together with the target-Hamiltonian spectrum, determine
the Fourier structure. Thus, the interplay between trainable
Hamiltonian spectra, open-system reduction, and the resulting Fourier
structure has not been explicitly characterized in this setting.

In this work, we bridge these two research directions through the
\emph{Interactive Quantum Classifier} (\IQC{}), a quantum classifier
in which an explicit Hamiltonian interaction between a target
subsystem and an ancillary environment induces a reduced quantum
channel, first introduced by
\cite{zhangInteractiveQuantumClassifier2021}. Unlike conventional
PQCs, where classical information is usually introduced through
feature-map unitaries, the \IQC{} encodes the input through an
input-dependent environmental Hamiltonian, whose spectral structure
determines the accessible frequencies of the resulting prediction
function. Predictions are obtained from measurements on the reduced
target state after tracing out the environment. This construction
provides an analytically tractable framework in which the reduced
dynamics and the resulting prediction function can be derived
explicitly. The \IQC{} was further modified by the \IQCail{} model in
\cite{britoQuantumClassifierBased2024}, which changed the input data
encoding to an amplitude encoding, and by the \IQCalpha{} model in
\cite{ferre2026interactive}, which introduced a systematic training of
different parameterizations of the target Hamiltonian for regression
problems.

Building upon previous formulations, we present a generalized
framework called \linebreak{} \IQCmulti{}, which recovers the previous
Hamiltonian-based \IQC{} instances as special cases. In this formulation,
the environment dimension becomes an independent architectural
parameter, allowing additional control over the accessible frequency
structure. Moreover, the resulting Fourier frequencies have a
multidimensional structure, allowing each frequency component to
involve linear combinations of multiple input features.

We evaluate all previously considered Hamiltonian-based \IQC{}
parameterizations on binary and multiclass classification benchmarks
and investigate how different forms of Hamiltonian freedom affect
predictive performance. The results show that increasing the available
degrees of freedom can improve classification performance, but does
not lead to a universal ordering between models. Instead, the observed
performance depends on the compatibility between the Fourier structure
accessible to each parameterization and the structure of the
underlying dataset.

Beyond predictive performance, we investigate the relationship between
Hamiltonian-induced Fourier structure and quantum expressibility using
the fidelity-distribution framework introduced by
\cite{sim2019expressibility}. Our results show that Haar-based
expressibility and classification performance capture different
aspects of the generated quantum models. In particular, architectures
that generate state ensembles closer to the Haar reference do not
necessarily achieve superior classification performance, highlighting
the distinction between global state-ensemble expressibility and
task-relevant Fourier expressivity.

The remainder of this paper is organized as
follows. Section~\ref{sec:background} reviews the Fourier
representation of parametrized quantum circuits and the Interactive
Quantum Classifier architecture. Section~\ref{sec:theory} derives the
reduced dynamics and Section~\ref{sec:fourier} establishes the Fourier
characterization of \IQC{}s. Section~\ref{sec:iqc-models} introduces the
Hamiltonian parameterizations motivated by this
framework. Section~\ref{sec:experiments} presents the experimental
methodology and Section~\ref{sec:results} the experimental results,
including classification and expressibility
analyses. Section~\ref{sec:discussion} discusses the implications of
the results, and Section~\ref{sec:conclusion} concludes the paper.

\section{Background}
\label{sec:background}

\subsection{Fourier representation of PQCs}

A fundamental result in quantum machine learning is the Fourier
characterization of parametrized quantum circuits (PQCs) introduced by
\cite{schuldEffectDataEncoding2021}. Consider a PQC with classical
input $\bm{x}\in\mathcal{X}\subset\mathbb{R}^{N_f}$, trainable
parameters $\bm{\theta}\in\mathbb{R}^{N_\theta}$, and unitary
evolution $U(\bm{\theta},\bm{x})$. For an initial state $\ket{0}$ and
a measurement operator $M$, the model output is
\begin{equation}
f(\bm{\theta},\bm{x})
=
\langle 0|
U^\dagger(\bm{\theta},\bm{x})
M
U(\bm{\theta},\bm{x})
|0\rangle,
\end{equation}
which can be expressed as a finite Fourier series in the input features,
\begin{equation}
\label{eq:fourier-exp-schuld}
f(\bm{\theta},\bm{x})
=
\sum_{\bm{\omega}\in\Omega}
c_{\bm{\omega}}(\bm{\theta})
e^{i\bm{\omega}\cdot\bm{x}},
\end{equation}
where $\Omega$ is the set of accessible frequencies and
$c_{\bm{\omega}}(\bm{\theta})$ are the corresponding Fourier
coefficients.

The derivation considers a layered PQC in which trainable gates are
interleaved with data-encoding operators,
\begin{equation}
U(\bm{\theta},\bm{x})
=
W^{(L+1)}(\bm{\theta})
\overleftarrow{\prod_{l=1}^{L}}
S(\bm{x})
W^{(l)}(\bm{\theta}),
\end{equation}
where
\begin{equation}
\label{eq:5}
\overleftarrow{\prod_{l=1}^{L}} A_l
\equiv
A_L A_{L-1}\cdots A_1
\end{equation}
denotes the right-ordered operator product. For a separable encoding,
the data-encoding operator can be written as
\begin{equation}
S(\bm{x})
=
e^{-ix_1H_1}
\otimes
\cdots
\otimes
e^{-ix_{N_f}H_{N_f}},
\end{equation}
where the Hermitian generators $H_j$ determine the accessible
frequencies. In particular, for each feature,
\begin{equation}
\Omega_j
=
\left\{
\lambda-\lambda'
:
\lambda,\lambda'\in\mathrm{spec}(H_j)
\right\},
\end{equation}
and, for the separable encoding above, the multidimensional frequency
set is
\begin{equation}
\Omega
=
\Omega_1
\times
\cdots
\times
\Omega_{N_f}.
\end{equation}
Thus, for fixed encoding Hamiltonians, the Hamiltonian spectra determine
the accessible Fourier frequencies, while the remaining circuit
parameters determine the corresponding Fourier coefficients.

The same structure can be expressed in density-matrix form as
\begin{equation}
\label{eq:schuld-objective-function-trace}
f(\bm{\theta},\bm{x})
=
\Tr
\left[
U(\bm{\theta},\bm{x})
\rho_0
U^\dagger(\bm{\theta},\bm{x})
M
\right],
\end{equation}
where $\rho_0$ is the initial state. More generally, the dependence
on the input can be introduced either through the unitary evolution or
through an input-dependent initial state. In the latter case, the model
can be written as
\begin{equation}
f(\bm{\theta},\bm{x})
=
\Tr
\left[
U(\bm{\theta})
\rho(\bm{x})
U^\dagger(\bm{\theta})
M
\right],
\end{equation}
where $\rho(\bm{x})$ denotes an input-dependent quantum state. These
two forms correspond to distinct data-loading strategies: encoding the
input in the Hamiltonian governing the evolution, or encoding it in
the initial quantum state. Both formulations are relevant to the
models considered in this work. In particular, the Hamiltonian-based
encoding leads to input-dependent Fourier frequencies, whereas the
amplitude-encoded initial state gives rise to input-dependent Fourier
coefficients.

\paragraph{Generalizations}

The Fourier framework has subsequently been extended in several
directions.  \cite{casasMultidimensionalFourierSeries2023a}
investigated the expressibility of multidimensional Fourier series
generated by quantum circuits, characterizing the degrees of freedom
required to represent general multivariate Fourier functions and their
scaling with circuit resources.  In a different direction,
\cite{jaderbergLetQuantumNeural2023} introduced trainable-frequency
quantum models in which parameters of the encoding generators are
optimized together with the variational circuit. Thus, the
optimization can modify not only the Fourier coefficients but also the
accessible frequency spectrum.

These results emphasize that the Hamiltonians used to encode classical
data play a fundamental role in determining the function class
accessible to a quantum learning model. This observation motivates the
analysis developed in the following sections, where we investigate how
the spectrum and eigenvalue gaps of an input-dependent interaction
Hamiltonian determine the Fourier structure of the reduced dynamics and
of the resulting classifier.

\section{Hamiltonian Formulation of Interactive Quantum Classifiers}
\label{sec:theory}

The Interactive Quantum Classifier (\IQC{}) introduced by
\cite{zhangInteractiveQuantumClassifier2021} can be formulated as a
Hamiltonian-induced reduced-channel quantum classifier. It was
subsequently generalized in \cite{britoQuantumClassifierBased2024},
which shifted the data encoding from the environment Hamiltonian to
the initial environment state, and in \cite{ferre2026interactive},
which investigated regression by training the coefficients of a
Pauli-linear target Hamiltonian together with the feature-wise
trainable environment Hamiltonian.

In this section, we introduce a generalized \IQC{} formulation, the
\IQCmulti{} model\footnote{The ``Multi'' suffix refers to the multidimensional Fourier frequencies generated by the model, whose frequency vectors can involve linear combinations of multiple input features.}, which
encompasses the Hamiltonian-based \IQC{} variants considered previously
as particular cases, while \IQCail{} remains distinct because it encodes
the input in the initial environment state. Rather than emphasizing a
particular circuit implementation, we describe the model directly
through its Hamiltonian evolution, which enables an analytical
characterization of both its reduced dynamics and its Fourier
representation.

The \IQCmulti{} classifier consists of a single target qubit interacting
with an ancillary environment under a joint Hamiltonian evolution.
The environment encodes the classical input through the eigenvalues
of its Hamiltonian, which in turn determine the conditional unitary
evolution of the target subsystem. After the interaction, the
environment is discarded and the prediction is obtained from an
observable
acting on the reduced target state. This reduced-channel formulation
provides the mathematical foundation for the Fourier analysis
developed in the next section.

\subsection{Hamiltonian formulation}

Consider the composite Hilbert space
\begin{equation}
\mathcal H=\mathcal H_t\otimes\mathcal H_e,
\end{equation}
where $\dim(\mathcal H_t)=2$ and $\dim(\mathcal H_e)=\envdim$. The
classical input is given by
$\bm{x}\in\mathcal X\subset\mathbb R^{N_f}$.

The trainable parameters are collected into
\begin{equation}
\label{eq:theta}
\bm\theta=(\alpha_0,\bm\alpha,W),
\end{equation}
where $\alpha_0\in\mathbb{R}$ and $\bm\alpha\in\mathbb{R}^3$
parametrize the target Hamiltonian, while
$W\in\mathbb{R}^{\envdim\times N_f}$ parametrizes the environment
Hamiltonian.

The joint evolution is
\begin{equation}
\label{eq:iqc-unitary}
U(\bm\theta,\bm x)
=
e^{-iH_t(\alpha_0,\bm\alpha)\otimes H_e(W,\bm x)}.
\end{equation}
The target Hamiltonian is written as
\begin{equation}
\label{eq:target-hamiltonian}
H_t(\alpha_0,\bm\alpha)
=
\alpha_0\mathbb{I}
+
\bm\alpha\cdot\bm\sigma,
\end{equation}
where
\begin{equation}
\bm\alpha=(\alpha_x,\alpha_y,\alpha_z),
\qquad
\bm\sigma=(\sigma_x,\sigma_y,\sigma_z).
\end{equation}
For the numerical implementation, we use the trace-normalized
Hamiltonian
\begin{equation}
\label{eq:normalized-target-hamiltonian}
\widetilde{H}_t
=
\frac{H_t}{\operatorname{Tr}(H_t)}
=
\frac{1}{2}\mathbb{I}
+
\frac{\bm\alpha}{2\alpha_0}\cdot\bm\sigma.
\end{equation}
The trace-normalized parametrization is used only for models with
$\alpha_0 \neq 0$. The original \IQC{}, defined by $\alpha_0=0$, is
instead treated using its unnormalized Hamiltonian
\eqref{eq:target-hamiltonian}.

The identity component in \eqref{eq:normalized-target-hamiltonian}
contributes only a global phase to the unitary evolution and therefore
does not affect the reduced state or the prediction function. We
henceforth omit this phase and consider the traceless component of the
target Hamiltonian. Meanwhile, the trace normalization fixes the
overall scale of the traceless part of the Hamiltonian used in the
numerical implementation, and this choice will be considered in the
analysis.

The environment Hamiltonian is assumed diagonal in the computational
basis,
\begin{equation}
H_e(W,\bm x)
=
\operatorname{diag}(W\bm x),
\end{equation}
where
\begin{equation}
\label{eq:w-matrix}  
W=
\begin{pmatrix}
\bm w_0^{T}\\
\vdots\\
\bm w_{\envdim-1}^{T}
\end{pmatrix}
\end{equation}
where
\begin{equation}
  \label{eq:env-eigenvalues}
  \lambda_l(\bm x)=\bm w_l^{T}\bm x = w_{l,0}x_{0} + w_{l,1}x_{1}+\dots+w_{l,N_{f}-1}x_{N_{f}-1}
\end{equation}
denotes the $l$-th environmental eigenvalue.  Therefore, $W$ maps the
input feature vector to the $\envdim$ eigenvalues of the environmental
Hamiltonian, with each row of $W$ defining a linear combination of
input features associated with one environmental eigenvalue. The
classical input is thus encoded directly into the eigenvalues of the
environment Hamiltonian.

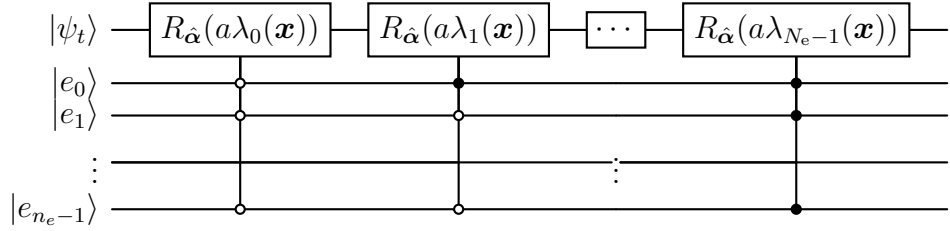
\begin{figure}[h]
\centering
\begin{quantikz}[row sep=0.28cm,column sep=0.5cm]
\lstick{$\ket{\psi_t}$}
&
\gate{R_{\hat{\bm{\alpha}}}(a\lambda_0(\bm{x}))}
&
\gate{R_{\hat{\bm{\alpha}}}(a\lambda_1(\bm{x}))}
&
\gate{\cdots}
&
\gate{R_{\hat{\bm{\alpha}}}(a\lambda_{\envdim-1}(\bm{x}))}
&
\\
\lstick{$\ket{e_0}$}
&
\octrl{-1}
&
\ctrl{-1}
&
&
\ctrl{-1}
&
\\
\lstick{$\ket{e_1}$}
&
\octrl{-2}
&
\octrl{-2}
&
&
\ctrl{-2}
&
\\
\lstick{$\vdots$}
&
\qw
&
\qw
&
\vdots
&
\qw
&
\\
\lstick{$\ket{e_{n_e-1}}$}
&
\octrl{-4}
&
\octrl{-4}
&
&
\ctrl{-4}
&
\end{quantikz}
\caption{Circuit representation of the \IQC{} architecture. The target
qubit is rotated by $R_{\hat{\bm{\alpha}}}(a\lambda_l(\bm{x}))$
depending on the state of the environment register.}
\label{fig:iqc-circuit}
\end{figure}

Since $H_e$ is diagonal, the joint evolution \eqref{eq:iqc-unitary}
admits the decomposition
\begin{equation}
\label{eq:iqc-multiplexor}
U(\bm\theta,\bm x)
=
\sum_{l=0}^{\envdim-1}
R_{\hat{\bm\alpha}}
\!\left(
\kappa\lambda_l(\bm x)
\right)
\otimes
|l\rangle\langle l|,
\end{equation}
where
\begin{equation}
\label{eq:kappa-parameter}
\kappa =
\left\{
  \begin{aligned}
           &\frac{\|\bm{\alpha}\|}{2\alpha_0},\quad &\alpha_{0}\neq 0 \quad &(\text{trace-normalized})\\[5pt]
           &\|\bm{\alpha}\|,\quad &\alpha_{0}= 0\quad &(\text{unnormalized}).
  \end{aligned}
  \right.
\end{equation}
and
\begin{equation}
  \label{eq:su2-rotation}
  R_{\hat{\bm\alpha}}(\phi)
  =
  e^{-i\phi\hat{\bm\alpha}\cdot\bm\sigma},\qquad \hat{\bm\alpha}
  =
  \frac{\bm\alpha}{\|\bm\alpha\|}.
\end{equation}
Eq.~(\ref{eq:iqc-multiplexor}) shows that the environment acts as a
quantum multiplexor when the environment dimension is
$\envdim=2^{n_e}$ (see Fig.~\ref{fig:iqc-circuit}), where each
computational basis state selects a different rotation of the target
qubit, while the corresponding rotation angle is determined by the
encoded input through the eigenvalues of the environment Hamiltonian.

\subsection{Reduced-channel representation}

Assuming an initial product state
\begin{equation}
\rho_0=\rho_t\otimes\rho_e,
\end{equation}
the reduced dynamics of the target qubit is
\begin{equation}
\label{eq:reduced-density-target}
\rho_t'(\bm\theta,\bm x)
=
\operatorname{Tr}_e
\!\left[
U(\bm\theta,\bm x)
\rho_0
U^\dagger(\bm\theta,\bm x)
\right]
=
\sum_{l=0}^{\envdim-1}
p_l
U_l
\rho_t
U_l^\dagger,
\end{equation}
where
\begin{equation}
  \label{eq:multiplexed-rotation}
U_l=
R_{\hat{\bm\alpha}}
\!\left(
\kappa\lambda_l(\bm x)
\right),
\end{equation}
and
\begin{equation}
  \label{eq:env-state-probabilites}
p_l=\langle l|\rho_e|l\rangle,
\end{equation}
with $p_l$ being the environment state probabilities. The prediction
function is thus obtained by measuring an observable $M$ on the
reduced target state,
\begin{equation}
\label{eq:prediction-function}
f(\bm\theta,\bm x)
=
\operatorname{Tr}
\!\left[
\rho_t'(\bm\theta,\bm x)
M
\right]
=
\sum_{l=0}^{\envdim-1}
p_l
f_l(\bm\theta,\bm x),
\end{equation}
where
\begin{equation}
\label{eq:fl-prediction-fct}
f_l(\bm\theta,\bm x)
=
\operatorname{Tr}
\!\left[
U_l
\rho_t
U_l^\dagger
M
\right].
\end{equation}

Eq.~(\ref{eq:reduced-density-target}) shows that the \IQC{} induces a
quantum channel on the target subsystem whose Kraus operators are the
multiplexed rotations $\sqrt{p_{l}}\,U_l$. The prediction function
\eqref{eq:prediction-function} is therefore a convex combination of
expectation-value quantum models, each associated with one eigenstate
of the environment Hamiltonian. This reduced-channel representation
forms the starting point for the Fourier characterization derived in
the next section.

\section{Fourier Representation of IQCs}
\label{sec:fourier}

In this section, we derive the analogue of the truncated Fourier
expansion in Eq.~\eqref{eq:fourier-exp-schuld} for the \IQC{} models
explicitly in terms of cosine and sine functions.  To this end, we
expand the prediction function in Eq.~\eqref{eq:fl-prediction-fct}
explicitly in terms of the input features.

For a single target qubit, the density matrix can be written in terms
of its Bloch vector $\bm{r}=(r_x,r_y,r_z)$ as
\begin{equation}
\rho_t
=
\frac{1}{2}
\left(
\mathbb{I}+\bm{r}\cdot\bm{\sigma}
\right).
\end{equation}
Although we consider pure initial states in the numerical experiments,
the following derivation applies to an arbitrary single-qubit state
for $\|\bm{r}\|\leq 1$.

To expand Eq.~\eqref{eq:fl-prediction-fct}, we first write the
multiplexed rotations \eqref{eq:multiplexed-rotation}  as
\begin{equation}
\label{eq:trigonometric-uq}
U_l
=
\cos(\theta_l)\mathbb{I}
-
i\sin(\theta_l)
(\hat{\bm\alpha}\cdot\bm{\sigma}),
\qquad
\theta_l=\kappa\lambda_l(\bm{x}).
\end{equation}
Using the Pauli-matrix identity
\begin{equation}
\label{eq:pauli-id-1}
[
\bm{r}\cdot\bm{\sigma},
\hat{\bm\alpha}\cdot\bm{\sigma}
]
=
2i
(\bm{r}\times\hat{\bm\alpha})\cdot\bm{\sigma},
\end{equation}
together with
\begin{equation}
\label{eq:puali-id-2}
(\hat{\bm\alpha}\cdot\bm{\sigma})
(\bm{r}\cdot\bm{\sigma})
(\hat{\bm\alpha}\cdot\bm{\sigma})
=
2(\hat{\bm\alpha}\cdot\bm{r})
(\hat{\bm\alpha}\cdot\bm{\sigma})
-
\bm{r}\cdot\bm{\sigma},
\end{equation}
we obtain
\begin{align}
U_l
(\bm{r}\cdot\bm{\sigma})
U_l^\dagger
&=
\cos^2(\theta_l)
(\bm{r}\cdot\bm{\sigma})
-
2\sin(\theta_l)\cos(\theta_l)
(\bm{r}\times\hat{\bm\alpha})\cdot\bm{\sigma}
\nonumber\\
&\quad
+
\sin^2(\theta_l)
\left[
2(\hat{\bm\alpha}\cdot\bm{r})
(\hat{\bm\alpha}\cdot\bm{\sigma})
-
\bm{r}\cdot\bm{\sigma}
\right]
\nonumber\\
&=
\bm{r}'_l\cdot\bm{\sigma},
\end{align}
where
\begin{equation}
\label{eq:rotated-bloch-vector-single}
\bm{r}'_l
=
\left[
\bm{r}
-
(\hat{\bm\alpha}\cdot\bm{r})
\hat{\bm\alpha}
\right]
\cos(2\theta_l)
+
(\hat{\bm\alpha}\times\bm{r})
\sin(2\theta_l)
+
(\hat{\bm\alpha}\cdot\bm{r})
\hat{\bm\alpha}.
\end{equation}
This is Rodrigues' rotation formula for a rotation of the Bloch vector
around the $\hat{\bm\alpha}$ axis by an angle
$2\theta_l$. Consequently, the reduced target state can be written as
\begin{equation}
\label{eq:rotated-target-density}
\rho_t'
=
\sum_{l=0}^{\envdim-1}
p_l
\frac{1}{2}
\left(
\mathbb{I}
+
\bm{r}'_l\cdot\bm{\sigma}
\right)
=
\frac{1}{2}
\left(
\mathbb{I}
+
\bm{r}'\cdot\bm{\sigma}
\right),
\end{equation}
where the reduced Bloch vector is
\begin{align}
\label{eq:rotated-bloch-vec}
\bm{r}'
&=
\sum_{l=0}^{\envdim-1}
p_l\bm{r}'_l
\nonumber\\
&=
\sum_{l=0}^{\envdim-1}
p_l
\Big\{
\left[
\bm{r}
-
(\hat{\bm\alpha}\cdot\bm{r})
\hat{\bm\alpha}
\right]
\cos(2\theta_l)
+
(\hat{\bm\alpha}\times\bm{r})
\sin(2\theta_l)
+
(\hat{\bm\alpha}\cdot\bm{r})
\hat{\bm\alpha}
\Big\}.
\end{align}
Equation~\eqref{eq:rotated-bloch-vec} shows directly that each component
of the reduced Bloch vector is a finite sum of sine and cosine functions
whose arguments are linear functions of the input features. The
prediction function therefore inherits the same finite Fourier
structure.

The prediction function in Eq.~\eqref{eq:prediction-function} is the
expectation value of a measurement operator $M$. A generic Hermitian
operator acting on a single qubit can be written as
\begin{equation}
\label{eq:9}
M
=
m_0\mathbb{I}
+
\bm{m}\cdot\bm{\sigma},
\end{equation}
where $m_0\in\mathbb{R}$ and
$\bm{m}\in\mathbb{R}^3$. Thus,
\begin{align}
\label{eq:f-in-M-terms}
f(\bm{\theta},\bm{x})
&=
\operatorname{Tr}
\left[
\rho_t'M
\right]
\nonumber\\
&=
m_0+\bm{m}\cdot\bm{r}'(\bm{\theta},\bm{x}).
\end{align}
Substituting Eq.~\eqref{eq:rotated-bloch-vec} into
Eq.~\eqref{eq:f-in-M-terms}, and using
$\theta_l=\kappa\bm{w}_l^T\bm{x}$, we obtain the finite Fourier-like
representation
\begin{align}
\label{eq:f-fourier-exp}
f(\bm{\theta},\bm{x})
&=
\sum_{l=0}^{\envdim-1}
\left[
A_l(\hat{\bm\alpha})
\cos\left(
2\kappa\bm{w}_l^T\bm{x}
\right)
+
B_l(\hat{\bm\alpha})
\sin\left(
2\kappa\bm{w}_l^T\bm{x}
\right)
\right]
+
C(\hat{\bm\alpha}),
\end{align}
where
\begin{equation}
\label{eq:f-fourier-coefs}
\begin{split}
A_l(\hat{\bm\alpha})
&=
p_l
\left[
\bm{m}\cdot\bm{r}
-
(\hat{\bm\alpha}\cdot\bm{r})
(\bm{m}\cdot\hat{\bm\alpha})
\right],
\\[5pt]
B_l(\hat{\bm\alpha})
&=
p_l
\bm{m}\cdot
(\hat{\bm\alpha}\times\bm{r}),
\\[5pt]
C(\hat{\bm\alpha})
&=
(\hat{\bm\alpha}\cdot\bm{r})
(\bm{m}\cdot\hat{\bm\alpha})
+
m_0.
\end{split}
\end{equation}
Since $\sum_l p_l=1$, the last term in
Eq.~\eqref{eq:f-fourier-coefs} is independent of the environment
index. The accessible frequencies of the prediction function are
therefore determined by the vectors
\begin{equation}
\bm{\omega}_l=2\kappa\bm{w}_l,
\qquad
l=0,\ldots,\envdim-1.
\end{equation}
In contrast, the coefficients $A_l$ and $B_l$ depend on the target
Hamiltonian direction ($\bhalpha$), the initial target state ($\br$),
the measurement operator $(\bm{m})$, and the environment probabilities
($p_{l}$).

This result establishes an explicit correspondence between the
Hamiltonian parameters of an \IQC{} and the Fourier structure of its
prediction function. In particular, the rows of the environmental
Hamiltonian parameter matrix $W$ determine the accessible frequency
vectors, while the target Hamiltonian and the measurement
configuration determine the corresponding Fourier coefficients. The
environment dimension therefore sets an upper bound on the number of
frequency components that can contribute to the prediction function.

Notice also that $m_0$ contributes an input-independent constant term
and can therefore be interpreted as a constant bias of the classifier.

The Fourier representation \eqref{eq:f-fourier-exp} is the main
theoretical result of this work. In the next section, we show how
previously proposed \IQC{} architectures and the new \IQCmulti{} model
arise as particular parameterizations of this general formulation, and
evaluate their empirical performance.

\section{IQC Models}
\label{sec:iqc-models}

The previous sections established a general Hamiltonian formulation for
interactive quantum classifiers together with their Fourier
representation. In this section, we specify the \IQC{} variants evaluated
throughout this work and describe the prediction function adopted in
the classification experiments.

\subsection{IQC Variants}
\label{sec:iqc-variants}

The Hamiltonian formulation introduced in Section~\ref{sec:theory}
naturally encompasses several \IQC{} architectures proposed in the
literature. These variants can be distinguished by the
parameterization of the target Hamiltonian, the environment
Hamiltonian, and, in the case of \IQCail{} (Amplitude Information
Loading), the strategy adopted to encode the classical input into the
environment subsystem. 

Accordingly, the original \IQC{} by
\cite{zhangInteractiveQuantumClassifier2021} is implemented using the
fixed parameter vector in \eqref{eq:target-hamiltonian}
\begin{equation}
 \alpha_{0}=0,\qquad \balpha=(1,1,1),
\end{equation}
whereas \IQCail{} follows the implementation proposed by
\cite{britoQuantumClassifierBased2024} and uses
\begin{equation}
\alpha_{0}=1,\qquad \balpha=(1,1,1).
\end{equation}
Because $\alpha_0$ is fixed, the normalization
\eqref{eq:normalized-target-hamiltonian} is not applied to these
models. The trainable models \IQCalpha{} and \IQCmulti{} optimize all
four parameters of the target Hamiltonian.

The original \IQC{} combines the fixed target Hamiltonian with the
feature-wise environment Hamiltonian
\begin{equation}
  \label{eq:iqc-env-hamiltonian}
H_e(\bx)=\operatorname{diag}(\bw\odot\bx),
\end{equation}
where
\begin{equation}
\bw\odot\bx=(w_1x_1,\ldots,w_{N_f}x_{N_f})
\end{equation}
is the Hadamard product.

\IQCail{}~\citep{britoQuantumClassifierBased2024} preserves the same
target Hamiltonian but transfers the data encoding from the
Hamiltonian to the initial environment state
\begin{equation}
  \label{eq:iqc-ail-env-state}
|\hat{\bx}\rangle
=
\sum_{l=0}^{N_{f}-1}
\frac{x_l}{\|\bx\|}|l\rangle,
\end{equation}
while the environment Hamiltonian becomes
\begin{equation}
  \label{eq:iqc-ail-env-hamiltonian}
H_e=\operatorname{diag}(\bw).
\end{equation}
For \IQCail{}, the environment dimension is set to \(N_e=N_f\), as
required by the amplitude encoding in
Eq.~\eqref{eq:iqc-ail-env-state}.

\IQCalpha{}~\citep{ferre2026interactive} preserves the original
feature-wise environment Hamiltonian \eqref{eq:iqc-env-hamiltonian}
while optimizing the four parameters of the target
Hamiltonian.

Finally, \IQCmulti{} replaces the feature-wise encoding by
\begin{equation}
\label{eq:iqc-multi-env-hamiltonian}
  H_e(\bx)=\operatorname{diag}(W\bx),
\end{equation}
with $W$ defined in \eqref{eq:w-matrix}, allowing each environment
eigenvalue to depend on arbitrary linear combinations of the input
features.

The \IQCmulti{} environment Hamiltonian
\eqref{eq:iqc-multi-env-hamiltonian} allows $N_{e}\neq N_{f}$ in
general, unlike the other models considered here.  In particular,
while the \IQCmulti{} Fourier frequency vectors in
\eqref{eq:f-fourier-exp} are multidimensional,
$\bw_{l}\in \mathbb{R}^{N_{f}}$, the other models have restricted
frequencies $\bw_{l}\propto \bm{e}_{l}$, where $\bm e_l$ denotes the
\(l\)-th unit basis vector. Therefore, we label the \IQCmulti{}
variants as \IQCmulti[$\envdim$]{} to distinguish different
environment dimensions.

Except for \IQCail{}, every \IQC{} architecture considered in this
work can be interpreted as a particular case of \IQCmulti{}. In
particular, \IQC{} and \IQCalpha{} are recovered by fixing the
environment dimension to $N_e=N_f$ and restricting
\begin{equation}
W=\operatorname{diag}(\bw)
\end{equation}
in \eqref{eq:iqc-multi-env-hamiltonian}, which yields
\eqref{eq:iqc-env-hamiltonian}. The original \IQC{} is further
obtained by fixing $\balpha=(1,1,1)$, whereas \IQCalpha{} optimizes
the target Hamiltonian. Consequently, \IQCmulti{} provides a unified
Hamiltonian formulation encompassing all \IQC{} architectures based on
Hamiltonian encoding considered in this work.  The experimental
comparison in the next section evaluates the effect of these different
Hamiltonian parameterizations, while also including \IQCail{} as a
distinct amplitude-encoding baseline.

Table~\ref{tab:model-variants} summarizes the
variants considered in this work, while Table~\ref{tab:notation}
summarizes the notation used throughout this work.

\begin{table}[!h]
\centering
\caption{\IQC{} variants evaluated in this work.}
\label{tab:model-variants}
\small
\setlength{\tabcolsep}{4pt}
\begin{tabular}{lccccc}
\toprule
Model & $H_t$ & $H_e$  & Parameters$^{\dagger}$ & $|\psi_e\rangle$ & Ref.\\
\midrule

\IQC{} & fixed &
$\operatorname{diag}(\bw\odot\bx)$
&
$N_f+1$
&
$|+\rangle$
&
\cite{zhangInteractiveQuantumClassifier2021}
\\

\IQCail{} & fixed &
$\operatorname{diag}(\bw)$
&
$N_f+1$
&
$|\hat{\bx}\rangle$
&
\cite{britoQuantumClassifierBased2024}
\\

\IQCalpha{} & $\alpha_0I+\balpha\cdot\bsigma$ &
$\operatorname{diag}(\bw\odot\bx)$
&
$N_f+5$
&
$|+\rangle$
&
\cite{ferre2026interactive}
\\

\IQCmulti{}-$N_e$ &$\alpha_0I+\balpha\cdot\bsigma$ &
$\operatorname{diag}(W\bx)$
&
$N_eN_f+5$
&
$|+\rangle$
&
This work
\\

\bottomrule
\end{tabular}
\begin{minipage}{\linewidth}
  \footnotesize \textbf{$^{\dagger}$Note:} The parameter count
  includes one trainable bias term and the four target Hamiltonian
  parameters, when applicable.
\end{minipage}
\end{table}
\vspace{-20pt}
\begin{table}[!htbp]
\centering
\caption{Notation used throughout the \IQC{} formulations.}
\label{tab:notation}
\begin{tabular}{ll}
\toprule
Symbol & Description\\
\midrule
$N_f$ & Number of input features\\
$N_e$ & Number of environmental eigenvalues / environment dimension\\
$n_e$ & Number of environmental qubits (if $N_{e}=2^{n_{e}}$)\\
$d$ & Total Hilbert-space dimension,
$d=2N_{e}$\\
$\bm{x}$ & Input feature vector\\
$W$ & Environmental encoding matrix,
$W\in\mathbb{R}^{N_e\times N_f}$\\
$\lambda_l(\bm{x})$ & $l$-th environmental eigenvalue\\
\bottomrule
\end{tabular}
\end{table}

\subsection{Prediction Function and Optimization}
\label{sec:prediction-function}

Throughout the experiments the target subsystem is initialized in the
state
\begin{equation}
|\psi_t\rangle=|+\rangle,
\end{equation}
the environment state is
\begin{equation}
  \label{eq:4}
  \ket{\psi_{e}} =\sum_{l=0}^{\envdim-1}\sqrt{p_{l}}\ket{l},\qquad \sum_{l=0}^{\envdim-1}p_{l}=1,
\end{equation}
and the prediction function is obtained by measuring
\begin{equation}
M=\sigma_z+bI,
\end{equation}
where $b$ denotes the trainable bias parameter. Substituting these
choices into the analytical expression \eqref{eq:f-fourier-exp}, we get
\begin{equation}
\label{eq:experimental-prediction}
f(\btheta,\bx)
=
\sum_{l=0}^{\envdim-1}
p_{l}\left[
-\hat\alpha_x\hat\alpha_z
\cos\!\left(2\kappa\bm w_l^{T}\bm x\right)
-
\hat\alpha_y
\sin\!\left(2\kappa\bm w_l^{T}\bm x\right)
\right]
+\hat\alpha_x\hat\alpha_z+b,
\end{equation}
where $\kappa$ is given by \eqref{eq:kappa-parameter}.
For \IQC{}, \IQCalpha{}, and \IQCmulti{}, we initialize the
environment in a uniform superposition, yielding
\(p_l=1/\envdim\).

For \(\IQCail{}\), the target Hamiltonian is unnormalized and
\(\boldsymbol\alpha=(1,1,1)\), so that \(\kappa=\sqrt3\).
Substituting $\bm w_l^{T}\bm x$ by $w_{l}$ and $p_{l}$ by
$x_{l}^{2}/\|\bx\|^{2}$ in \eqref{eq:experimental-prediction}, we get
the \IQCail{} prediction function
\begin{equation}
\label{eq:experimental-ail-prediction}
g(\btheta,\bx)
=
\sum_{l=0}^{\envdim-1}
\frac{x_{l}^{2}}{/\|\bx\|^{2}}\left[
-\frac{1}{3}
\cos\!\left(2\sqrt{3}w_l\right)
-
\frac{1}{\sqrt{3}}
\sin\!\left(2\sqrt{3}w_l\right)
\right]
+\frac{1}{3}+b.
\end{equation}
This can be rewritten as
\begin{equation}
  \label{eq:exp-ail-prediction-2}
  g(\btheta,\bx) = \hat{\bx}^{T}\tilde{W}\hat\bx +b +\frac{1}{3},
\end{equation}
where \(\hat{\bx}=\bx/\|\bx\|\) is the unit vector in the direction of
\(\bx\) and $\tilde{W} = \operatorname{diag}(\tilde{\bw})$ the
diagonal matrix with components
\begin{equation}
  \label{eq:wl-ail}
  \tilde{w}_{l} = -\frac{1}{3}
\cos\!\left(2\sqrt{3}w_l\right)
-
\frac{1}{\sqrt{3}}
\sin\!\left(2\sqrt{3}w_l\right) = -\frac{2}{3}\cos(2\sqrt{3}w_{l}-\frac{\pi}{3}). 
\end{equation}
Therefore, while the prediction functions of the other models have a
Fourier-like structure in the feature space with trainable
coefficients, with \IQCmulti{} having more general Fourier frequencies
than the other models, \IQCail{} actually has a homogeneous quadratic
decision boundary after normalization, constrained by the diagonal
structure of $\tilde{W}$ and by $|\tilde{w}_{l}| \leq 2/3$.

\section{Experimental Methodology}
\label{sec:experiments}

This section describes the experimental protocol used to evaluate the
six \(\IQC{}\)-based model variants introduced in
Section~\ref{sec:iqc-models} (see Table~\ref{tab:model-variants}). The
same preprocessing and evaluation framework was used for all models,
while two alternative optimization procedures were considered.

\subsection{IQC Settings}
\label{sec:iqc-settings}

All \IQC{} variants considered in this work use the same underlying
open-quantum-system-inspired binary classification framework and differ
only in the Hamiltonian and input-encoding parametrizations described in
Section~\ref{sec:iqc-models}.

During training, class labels were encoded as $\{-1,+1\}$ and the
trainable parameters were initialized independently from a uniform
distribution over $[-\pi,\pi]$. The mean squared error (MSE) between the
model output and the encoded labels was used as the training objective.

\subsection{Experimental Protocol}
\label{sec:experimental-protocol}

All experiments were implemented in Python version 3.11.5. The
numerical formulation of the models was implemented using
\texttt{NumPy} version 2.4.6 \citep{harris2020array} and
\texttt{jax.numpy} from \texttt{JAX} version 0.6.2
\citep{bradbury2018jax}. Experiments were conducted on a workstation
equipped with a 12th-generation Intel Core i7-12700H processor, 16GiB
of system memory, and an NVIDIA GeForce RTX 3060 Laptop GPU. JAX-based
computations, including automatic differentiation and model
optimization, were executed using the GPU backend. All quantum
computations reported in this work were performed through classical
numerical simulation.

Six \IQC{} model variants were evaluated on seven
binary-classification benchmarks and four real-world benchmarks,
comprising two binary and two multiclass datasets. For each dataset
and model, ten independent train/test partitions were generated using
distinct random seeds.

For the binary classification datasets, each repetition used an
$80\%/20\%$ train/test split. For the multiclass datasets, the same
repeated hold-out scheme was used, with class-stratified sampling to
preserve the class proportions in the training and test subsets.

For every partition, input features were rescaled to the $[0,1]$
interval using a min--max scaler fitted exclusively on the training
subset and subsequently applied, without refitting, to the corresponding
test subset. This procedure prevents information from the test set from
entering the preprocessing stage.

\subsection{Datasets}
\label{sec:datasets}

Two families of datasets were considered: synthetic benchmarks for
binary classification and real-world benchmarks comprising both binary
and multiclass classification problems.

\paragraph{Synthetic classification datasets.}
Seven synthetic benchmarks were used, all generated using utilities provided by \texttt{scikit-learn} version 1.9.0, a Python library for machine learning \citep{pedregosa2011scikit}:

\begin{itemize}
    \item \textbf{Blobs (2D, 4D, and 8D).}
    Synthetic isotropic Gaussian clusters generated with
    \texttt{sklearn.datasets.make\_blobs}, using $n=300$ samples, two
    centers, and a cluster standard deviation of $1.0$. Three variants
    with $2$, $4$, and $8$ input features were considered.

    \item \textbf{Circles.}
    A two-dimensional nonlinear classification dataset consisting of
    two concentric circles, generated with
    \texttt{sklearn.datasets.make\_circles} using $n=1000$ samples and
    no additive noise.

    \item \textbf{Moons.}
    A two-dimensional nonlinear classification dataset consisting of
    two interleaving half-moon-shaped classes, generated with
    \texttt{sklearn.datasets.make\_moons} using $n=1000$ samples and
    no additive noise.

    \item \textbf{Stripes.} A two-dimensional nonlinear classification dataset consisting of three parallel Gaussian stripes, generated from four isotropic Gaussian clusters using \texttt{sklearn.datasets.make\_blobs} with $n=400$ samples, four centers, a cluster standard deviation of $1.0$, and \texttt{random\_state=170}. An anisotropic linear transformation was applied to the generated samples, after which two clusters were superimposed to form the central stripe. The two superimposed clusters were assigned to one class, while the two outer stripes were assigned to the other class, resulting in two outer stripes surrounding a central stripe. 
    
    \item \textbf{XOR (exclusive OR).} A two-dimensional nonlinear classification dataset consisting of four isotropic Gaussian clusters arranged according to the XOR pattern, generated with \texttt{sklearn.datasets.make\_blobs} using $n=400$ samples, centers at $(-1,-1)$, $(-1,1)$, $(1,1)$, and $(1,-1)$, a cluster standard deviation of $0.25$, and \texttt{random\_state=0}. Clusters located at diagonally opposite positions were assigned to the same class, resulting in the characteristic exclusive-OR decision structure.
\end{itemize}

\paragraph{Real-world classification datasets.}
Four real-world benchmarks were used, obtained either from datasets
provided by the \texttt{scikit-learn} Python library
\citep{pedregosa2011scikit} or from publicly available data
repositories:

\begin{itemize}
    \item \textbf{Iris.}
    The classical Iris flower dataset
(\texttt{sklearn.datasets.load\_iris}) comprises 150 instances from three classes (\textit{Iris setosa}, \textit{Iris versicolor}, and \textit{Iris virginica}), described by four numerical features: \textit{sepal length}, \textit{sepal width}, \textit{petal length}, and \textit{petal width}.

\item \textbf{Wine.}  The Wine recognition dataset
  (\texttt{sklearn.datasets.load\_wine}) comprises 178 wine samples
   from three different cultivars grown in the same region of
  Italy, which define the three classification classes. Each instance
  is described by thirteen numerical features obtained from chemical
  analyses: \textit{alcohol}, \textit{malic acid}, \textit{ash},
  \textit{alcalinity of ash}, \textit{magnesium}, \textit{total
    phenols}, \textit{flavanoids}, \textit{nonflavanoid phenols},
  \textit{proanthocyanins}, \textit{color intensity}, \textit{hue},
  \textit{OD280/OD315 of diluted wines}, and \textit{proline}.

\item \textbf{Pima Diabetes.}
  The Pima Indians Diabetes dataset \citep{smith1988adap} was obtained through the OpenML repository (dataset ID 37) and comprises 768 instances described by eight clinical attributes: \textit{Number of pregnancies}, \textit{Plasma glucose concentration}, \textit{Diastolic blood pressure}, \textit{Triceps skinfold thickness}, \textit{2-hour serum insulin}, \textit{Body mass index (BMI)}, \textit{Diabetes pedigree function}, and \textit{Age}. The binary target indicates the presence or absence of diabetes.
  
\item \textbf{Caesarian Section.}
The Caesarian Section Classification dataset \citep{amin2018caesarian} was obtained from the UCI Machine Learning Repository and comprises 80 instances described by five attributes: \textit{Age}, \textit{Delivery number}, \textit{Delivery time}, \textit{Blood Pressure}, and \textit{Heart Problem}. The binary target attribute, \textit{Caesarian}, indicates whether a Caesarian section was performed. 

\end{itemize}

Because the underlying \IQC{} estimator supports binary
classification, all multiclass datasets were handled through a
one-vs-rest decomposition using \linebreak{}
\texttt{sklearn.multiclass.OneVsRestClassifier}, with one
independently trained binary \IQC{} instance for each class.

\subsection{Optimization Settings}
\label{sec:optimization-settings}

The \IQC{} estimator supports two alternative training procedures: a
gradient-based optimizer and a particle-swarm optimizer (PSO). Both
procedures were applied to the binary and multiclass datasets.

\paragraph{Gradient-based optimizer.}
Parameters were updated using the Adam optimizer \linebreak{}\citep{kingma2015adam}, as implemented in \texttt{Optax} version 0.2.8 \citep{hessel2020optax}, with gradients of the MSE loss computed via automatic differentiation using \texttt{JAX} version 0.6.2 (\texttt{jaxlib} version 0.6.2) \citep{bradbury2018jax}. Training was performed for $1000$ optimization steps.

The learning rate was evaluated at three values,
\[
\eta\in\{0.1,0.01,0.001\},
\]
with each value combined with the ten random seeds used in the repeated
hold-out protocol. No early-stopping criterion was used for this
optimizer; training therefore always ran for the full number of
optimization steps.

\paragraph{Particle-swarm optimizer (PSO).}
Parameters were alternatively optimized using the global-best particle
swarm optimizer, as implemented by
\texttt{pyswarms.single.GlobalBestPSO} in \texttt{PySwarms} version
1.3.0 \citep{miranda2018pyswarms}. The same MSE objective was
evaluated over the complete training partition.

The swarm consisted of $30$ particles, with inertia and acceleration
coefficients
\[
w=0.9,\qquad c_1=0.5,\qquad c_2=0.3.
\]
Optimization proceeded for up to $500$ iterations, with early stopping
when the training loss fell below $10^{-3}$. Since the learning rate is
not a parameter of PSO, this optimizer was executed once per random seed
rather than being repeated over the learning-rate sweep used for Adam.

For the multiclass experiments, one binary \IQC{} instance, with its own
independently optimized parameters, was trained per class under the
one-vs-rest scheme.

\subsection{Evaluation Metrics}
\label{sec:evaluation-metrics}

\paragraph{Classification.}
For all classification experiments, four performance metrics were recorded: accuracy, macro-averaged $F_1$-score, macro-averaged precision, and macro-averaged recall \citep{powers2011evaluation}.
The reported classification results are summarized as mean $\pm$
standard deviation over the ten repeated train/test partitions.

\paragraph{Expressibility.}
The expressibility of the quantum models was evaluated using the
fidelity-distribution framework introduced by 
\cite{sim2019expressibility}. For each model, we generated an ensemble
of global pure states by independently sampling both the classical
input and the model parameters. Each input feature was sampled
independently from a uniform distribution over $[0,1)$. Model-specific
weight parameters were also sampled independently from the same
distribution, including the bias parameter. For \IQCalpha{} and \IQCmulti{}, the components of the
target Hamiltonian parameter vector $\bm{\alpha}$ were additionally
sampled independently from $[0,1)$, whereas the target Hamiltonian was
kept fixed for \IQC{} and \IQCail{}.

For each model, $10\,000$ independent pairs of global output states
$\ket{\Psi_i}$ and $\ket{\Psi_j}$ were generated. The states include
both the target and environment subsystems and were obtained from the
joint unitary evolution applied to the corresponding initial product
state (see Sec.~\ref{sec:theory}). The pairwise state fidelity was
computed as
\begin{equation}
F_{ij}
=
\left|
\braket{\Psi_i|\Psi_j}
\right|^2.
\end{equation}
The resulting empirical fidelity distribution was estimated using 75
uniformly spaced bins over $[0,1]$.

For a $d$-dimensional Hilbert space, the fidelity distribution for
pairs of independent Haar-random pure states is
\citep{zyczkowskiAverageFidelityRandom2005}
\begin{equation}
p_{\mathrm{Haar}}(F)
=
(d-1)(1-F)^{d-2},
\qquad 0\leq F\leq1.
\end{equation}
Rather than estimating this reference numerically, its probability
within each histogram bin $[F_k,F_{k+1}]$ was computed analytically as
\begin{equation}
P_{\mathrm{Haar}}^{(k)}
=
(1-F_k)^{d-1}
-
(1-F_{k+1})^{d-1},\qquad k=0,\dots,N_{\text{bins}}-1,
\end{equation}
where $N_{\text{bins}}$ is the total number of bins, $F_0=0$ and $F_{N_{\text{bins}}}=1$. The discretized empirical and Haar fidelity distributions were then
compared using the Kullback--Leibler divergence
\citep{sim2019expressibility}
\begin{equation}
\mathrm{Expr}
=
D_{\mathrm{KL}}
\left(
P_{\mathrm{model}}\middle\|P_{\mathrm{Haar}}
\right)
=
\sum_{k=0}^{N_{\mathrm{bins}}-1}
P_{\mathrm{model}}^{(k)}
\log
\frac{P_{\mathrm{model}}^{(k)}}
{P_{\mathrm{Haar}}^{(k)}} ,
\qquad
N_{\mathrm{bins}}=75.
\end{equation}
Lower values of this divergence indicate a fidelity distribution
closer to the Haar reference and therefore greater expressibility
according to this metric. The Hilbert-space dimension $d$ was computed
for the full target--environment state as
\begin{equation}
d=2N_e,
\end{equation}
so that the reference distribution was determined separately for each
model.

Because the inputs and model parameters are independently resampled
for each realization, the resulting fidelity distribution
characterizes the state ensemble induced by each \IQC{} architecture under
the specified input and parameter sampling distributions. It therefore
measures a structural property of the model family rather than the
state ensemble produced by a particular set of trained parameters.

Our expressibility measure therefore characterizes the statistical
structure of the generated state ensemble of an architecture, whereas
classification performance reflects the subset of this space that can
be reached through training and provides a useful representation for a
particular dataset. This interplay is examined in the next section
through the empirical results.

\paragraph{Relative Expressibility}

The expressibility measure based on the KL divergence quantifies the
distance between the fidelity distribution generated by a model and the
corresponding Haar-random reference distribution. Therefore, smaller
values of $\mathrm{Expr}$ indicate that the generated state ensemble is
closer to the Haar ensemble for a fixed Hilbert-space dimension. However,
direct comparisons of $\mathrm{Expr}$ across different dimensions are
not straightforward, since the Haar reference distribution itself
depends on the dimension of the Hilbert space.

To enable comparisons between models with different global dimensions,
the expressibility can be normalized with respect to the corresponding
idle circuit, i.e., a circuit implementing only the identity operation
\citep{rasmussenReducingAmountSinglequbit2020}. For the \IQC{},
\IQCalpha{}, and \IQCmulti{} architectures, the initial state is
independent of the input. Consequently, the idle circuit produces the
same global state for every realization, and all pairwise fidelities are
equal to one. Since the empirical fidelity distribution is therefore
concentrated in the last histogram bin, its KL divergence with respect
to the Haar distribution can be obtained analytically. For uniformly
spaced bins over $[0,1]$, the Haar probability of the last bin is
$N_{\mathrm{bins}}^{-(d-1)}$, yielding
\begin{equation}
\label{eq:idle_expr}
\mathrm{Expr}_{\mathrm{idle}}
=
(d-1)\log(N_{\mathrm{bins}}),
\end{equation}
where $d$ is the global Hilbert-space dimension.

The \IQCail{} architecture requires a different treatment because its
initial environment state depends explicitly on the input through
amplitude encoding. Thus, even under the idle circuit, different inputs
generate different states,
\begin{equation}
\ket{\Psi_{\mathrm{idle}}(\bx)}
=
\ket{+}\otimes\ket{\hat{\bx}},
\qquad
\ket{\hat{\bx}}
=
\frac{1}{\|\bx\|}
\sum_l x_l\ket{l}.
\end{equation}
The corresponding idle fidelities are therefore
\begin{equation}
F_{ij}^{\mathrm{idle}}
=
\left|
\braket{\hat{\bx}_i}{\hat{\bx}_j}
\right|^2,
\end{equation}
and the \IQCail{} idle expressibility is obtained numerically by
constructing the empirical fidelity distribution from independently
sampled input pairs and computing its KL divergence with respect to the
corresponding Haar distribution.

The relative expressibility is then defined as
\begin{equation}
\label{eq:relative_expr}
\mathrm{RExpr}
=
-\ln
\left(
\frac{\mathrm{Expr}}
{\mathrm{Expr}_{\mathrm{idle}}}
\right).
\end{equation}
Higher values of $\mathrm{RExpr}$ indicate a larger reduction of the
KL divergence relative to the corresponding idle ensemble, and
therefore a larger approach toward the Haar reference induced by the
model dynamics. This normalization enables comparisons of
expressibility across \IQC{} parameterizations with different
Hilbert-space dimensions by accounting for the corresponding idle
baseline and their distinct input-encoding mechanisms.

\section{Results}
\label{sec:results}

\subsection{Overall performance}
\label{sec:overall-performance}

The classification results for the \IQC{} architectures are provided
in Tables~\ref{tab:results_binary} and~\ref{tab:results_realworld} in
Appendix \ref{sec:class-results}. The reported values correspond to
the mean performance of the best training configuration selected for
each model and dataset, where the best configuration is determined by
its mean performance across the repeated train/test partitions. The
corresponding standard deviations are computed over these repeated
partitions, as described in Sec.~\ref{sec:experimental-protocol}.

\begin{figure}[!h]
    \centering
    \includegraphics[width=\linewidth]{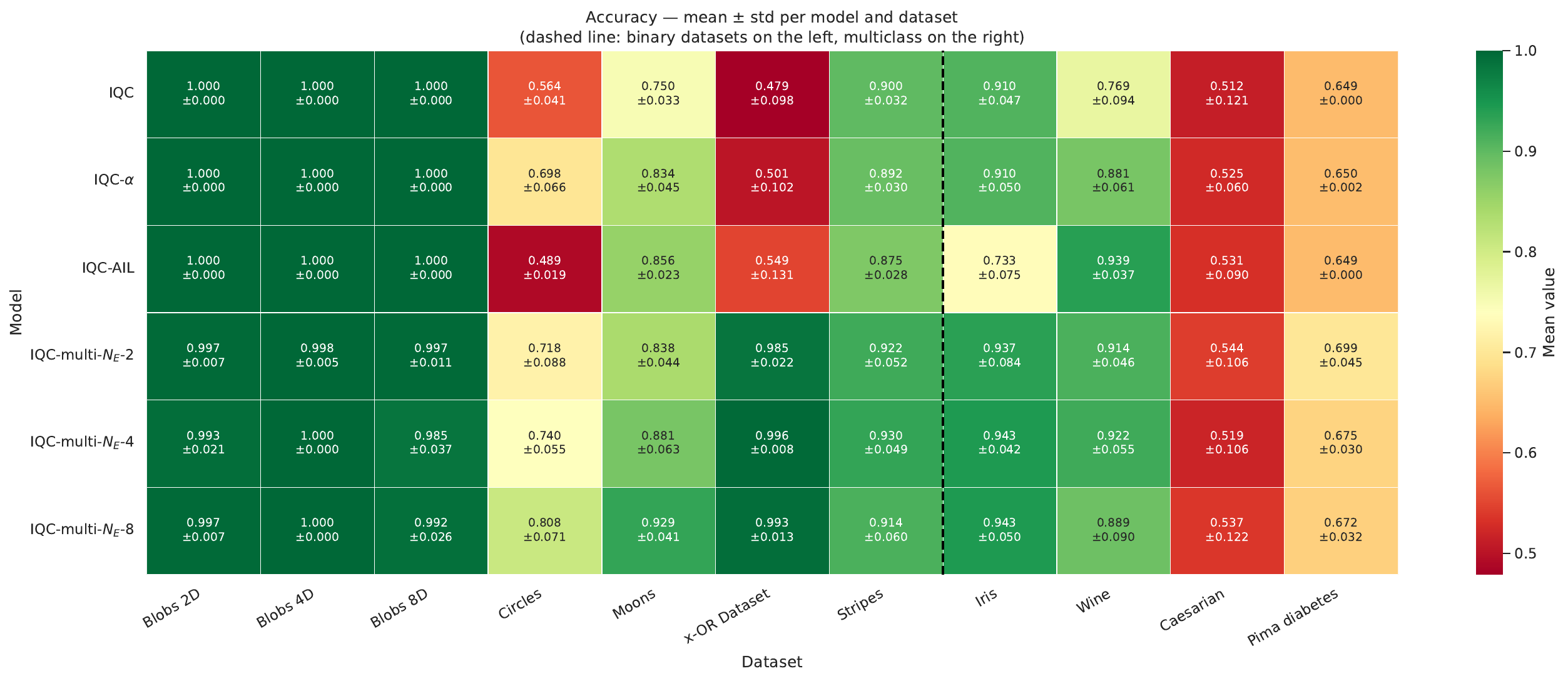}
    \caption{Accuracy (mean $\pm$ standard deviation) obtained by each
    \IQC{} variant across the eleven benchmark datasets. The dashed
    vertical line separates binary datasets (left) from multiclass
    datasets (right).}
    \label{fig:heatmap_acc}
\end{figure}
\begin{figure}[!h]
    \centering
    \includegraphics[width=\linewidth]{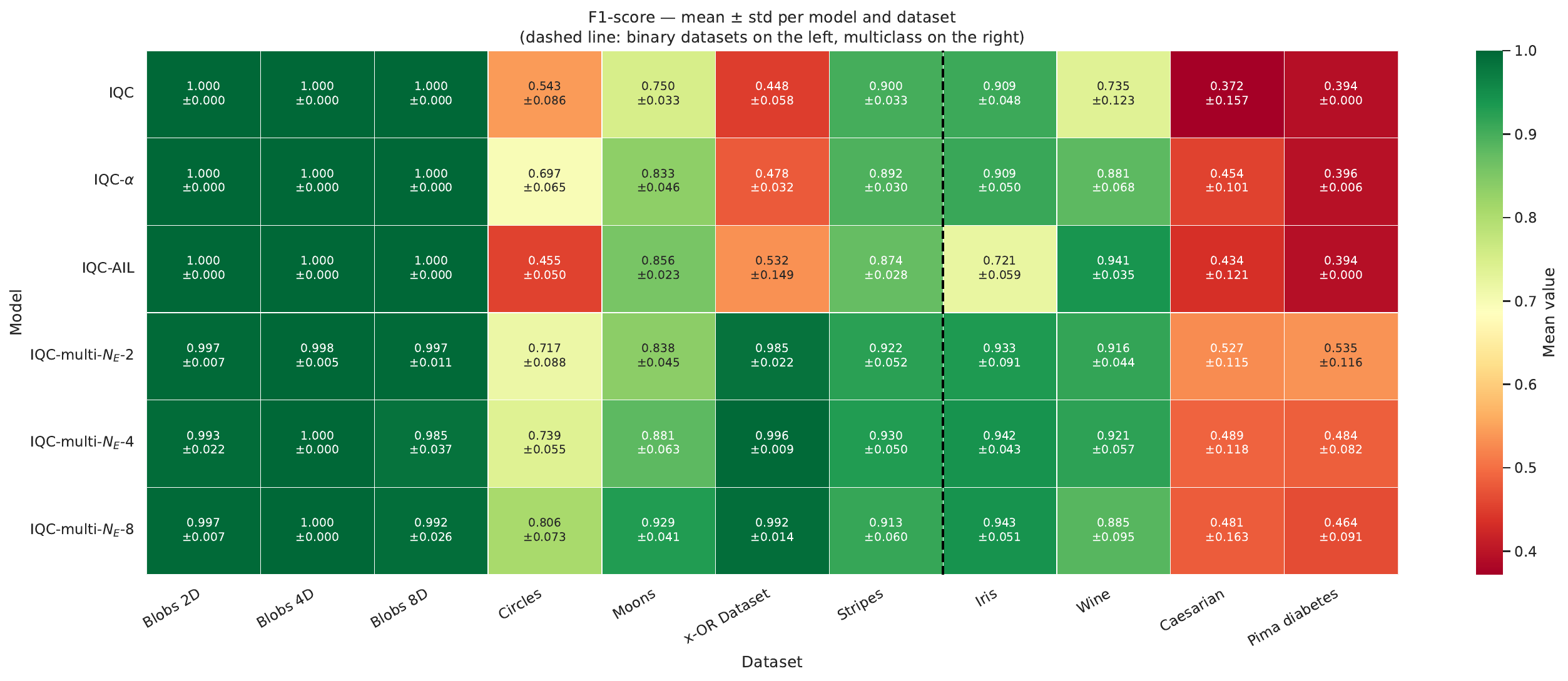}
    \caption{F1-score (mean $\pm$ standard deviation) obtained by each
    \IQC{} variant across the eleven benchmark datasets. The dashed
    vertical line separates binary datasets (left) from multiclass
    datasets (right).}
    \label{fig:heatmap_f1score}
\end{figure}

Figures~\ref{fig:heatmap_acc} and~\ref{fig:heatmap_f1score} provide a
compact visualization of the best accuracy and F1-score results across
all datasets and models. The heatmaps highlight both the strong
performance of the multidimensional variants on several of the more
challenging synthetic benchmarks and the dataset-dependent nature of
the model ranking.

The three Blobs datasets exhibit near-saturated performance, with all
models reaching accuracy values close to one, whereas the more
challenging synthetic datasets provide substantially greater
separation between the models. On Moons, for example, \IQCmulti[8]
achieves an accuracy of $0.929$, compared with $0.881$ for
\IQCmulti[4] and $0.856$ for \IQCail{}. On XOR, the multidimensional
variants produce a much larger separation: \IQCmulti[4] reaches an
accuracy of $0.996$, while the non-multidimensional variants remain
close to chance level.

The real-world datasets further illustrate the dataset dependence of
the ranking. On Iris, \IQCmulti[4] and \IQCmulti[8] both achieve an
accuracy of $0.943$. On Pima Diabetes, \IQCmulti[2] obtains the
highest accuracy, $0.699$. In contrast, Wine provides a clear
counterexample to a monotonic relationship between environment size
and predictive performance: \IQCail{} achieves the highest accuracy
and F1-score, $0.939$ and $0.941$, respectively.

A marked divergence between accuracy and F1-score is observed on
the Caesarian and Pima Diabetes datasets. For these datasets, accuracy
remains substantially higher than F1-score, illustrating why accuracy
alone may be insufficient to characterize classifier performance in
the presence of class imbalance.

\begin{table}[!ht]
\centering
\caption{Overall classification performance averaged across the eleven
  benchmark datasets after selecting the best training configuration
  for each model--dataset pair.}
\label{tab:overall-performance}
\begin{tabular}{lcccc}
\toprule
Model & Accuracy & F1-score & Precision & Recall\\
\midrule
\IQC{} & 0.7951 & 0.7448 & 0.7509 & 0.7765\\
\IQCail{} & 0.7998 & 0.7555 & 0.7633 & 0.7843\\
  \IQCalpha{} & 0.8331 & 0.7967 & 0.8188 & 0.8220\\
\IQCmulti[2] & 0.8490 & 0.8287 & 0.8507 & 0.8396\\
\IQCmulti[4] & 0.8509 & 0.8261 & 0.8576 & 0.8406\\
\IQCmulti[8] & \textbf{0.8630} & \textbf{0.8329} & \textbf{0.8601} & \textbf{0.8522}\\
\bottomrule
\end{tabular}
\label{tab:overall-classification}
\end{table}

For an aggregate comparison across the eleven datasets, the results
are summarized in Table~\ref{tab:overall-performance}.  Across the
eleven datasets, \IQCmulti[8] achieves the highest mean value for all
four metrics, with accuracy, F1-score, precision, and recall of
$0.863$, $0.833$, $0.860$, and $0.852$, respectively. The aggregate
ranking does not imply that the same model is optimal for every
dataset, as illustrated by the dataset-level results discussed
above. The \IQCmulti[4] and \IQCmulti[2] variants also achieve high
aggregate performance, while \IQCalpha{}, \IQCail{}, and \IQC{}
exhibit lower aggregate values.  A potential explanation for these
results is presented in Sec.~\ref{sec:discussion}.

\subsection{Statistical Analysis}
\label{sec:statistical-analysis}

To assess whether the performance differences among the evaluated
models were statistically significant, the Friedman test
\citep{friedman1937} was applied independently to each dataset--metric
combination, using the ten repeated train/test partitions as paired
observations. The test was performed using the
\texttt{friedmanchisquare} implementation provided by \texttt{SciPy}
version 1.17.1 \citep{virtanen2020scipy}, with a significance level of
$\alpha=0.05$. Significant differences among the models were detected
for 7 datasets in accuracy, 7 in $F_1$-score, 6 in precision, and 7 in
recall. No significant differences were detected for the three Blobs
datasets or for the Caesarian Section dataset across the four
evaluation metrics.

For dataset--metric combinations exhibiting a significant Friedman
test, pairwise comparisons between models were performed using the
two-sided Wilcoxon signed-rank test~\citep{wilcoxon1945}, with the ten
repeated train/test partitions treated as paired observations. The
test was implemented using \texttt{SciPy} version
1.17.1~\citep{virtanen2020scipy}. To account for multiple comparisons,
for each dataset--metric combination, the resulting pairwise
$p$-values were adjusted using the Holm procedure~\citep{holm1979}, as
implemented in \texttt{statsmodels} version 0.14.6, with a
significance level of $\alpha=0.05$. For the $F_1$-score, the overall
post-hoc results are summarized in
Table~\ref{tab:wilcoxon-holm-f1-overall}.

\begin{table}[!h]
  \centering
  \caption{Aggregate counts of statistically significant wins, losses,
    and non-significant pairwise comparisons for the F1-score across
    the seven datasets with a significant Friedman test.}
\small
\label{tab:wilcoxon-holm-f1-overall}
\begin{tabular}{lrrrr}
  \toprule
  Model & Wins & Losses & No difference & Balance$^{\dagger}$ \\
  \midrule
  \IQCmulti[8] & 10 & 0 & 25 & $+10$ \\
  \IQCmulti[4] & 8  & 0 & 27 & $+8$ \\
  \IQCmulti[2] & 7  & 1 & 27 & $+6$ \\
  \IQCalpha{}      & 3  & 4 & 28 & $-1$ \\
  \IQCail{}           & 2  & 14 & 19 & $-12$ \\
  \IQC{}               & 1  & 12 & 22 & $-11$ \\
  \bottomrule
\end{tabular}
\begin{minipage}{\linewidth}
  \footnotesize \textbf{$^{\dagger}$Note:} A significant win (loss)
  denotes a pairwise comparison with a Holm-adjusted \(p\)-value below
  0.05 in which the model in the corresponding row achieved a higher
  (lower) F1-score than the model in the comparison column.
\end{minipage}
\end{table}

These results are consistent with the higher aggregate performance of
the multidimensional variants, while also showing that their advantage
is not universal. In particular, most pairwise comparisons for
\IQCmulti[8] are not statistically significant, despite its generally
high mean performance across datasets. Conversely, individual datasets
can favor other parametrizations, as illustrated by the strong
performance of \IQCail{} on the Wine dataset. The complete Wilcoxon
signed-rank results with Holm correction for the individual
dataset--metric combinations are reported in the Appendix
\ref{app:wilcoxon-holm}.

\subsection{Decision regions}
\label{sec:decision-regions}

\begin{figure}[!htbp]
    \centering
    \includegraphics[width=\linewidth]{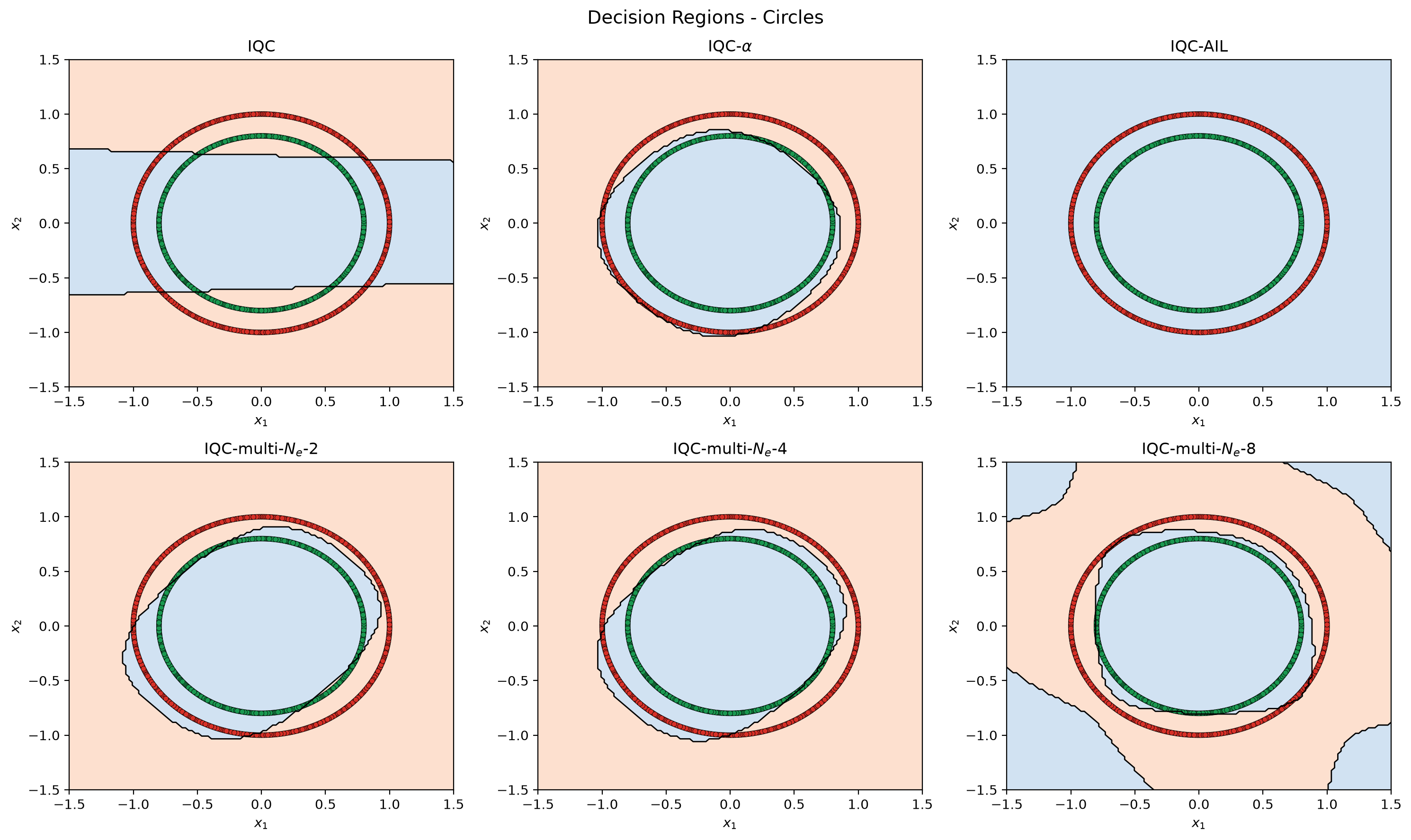}
    \caption{Decision regions learned by the six \IQC{} variants on the
    Circles dataset. The plots show the resulting classification regions
    together with the data points for the two classes.}
    \label{fig:decision_regions_circle}
\end{figure}

\begin{figure}[!htbp]
    \centering
    \includegraphics[width=\linewidth]{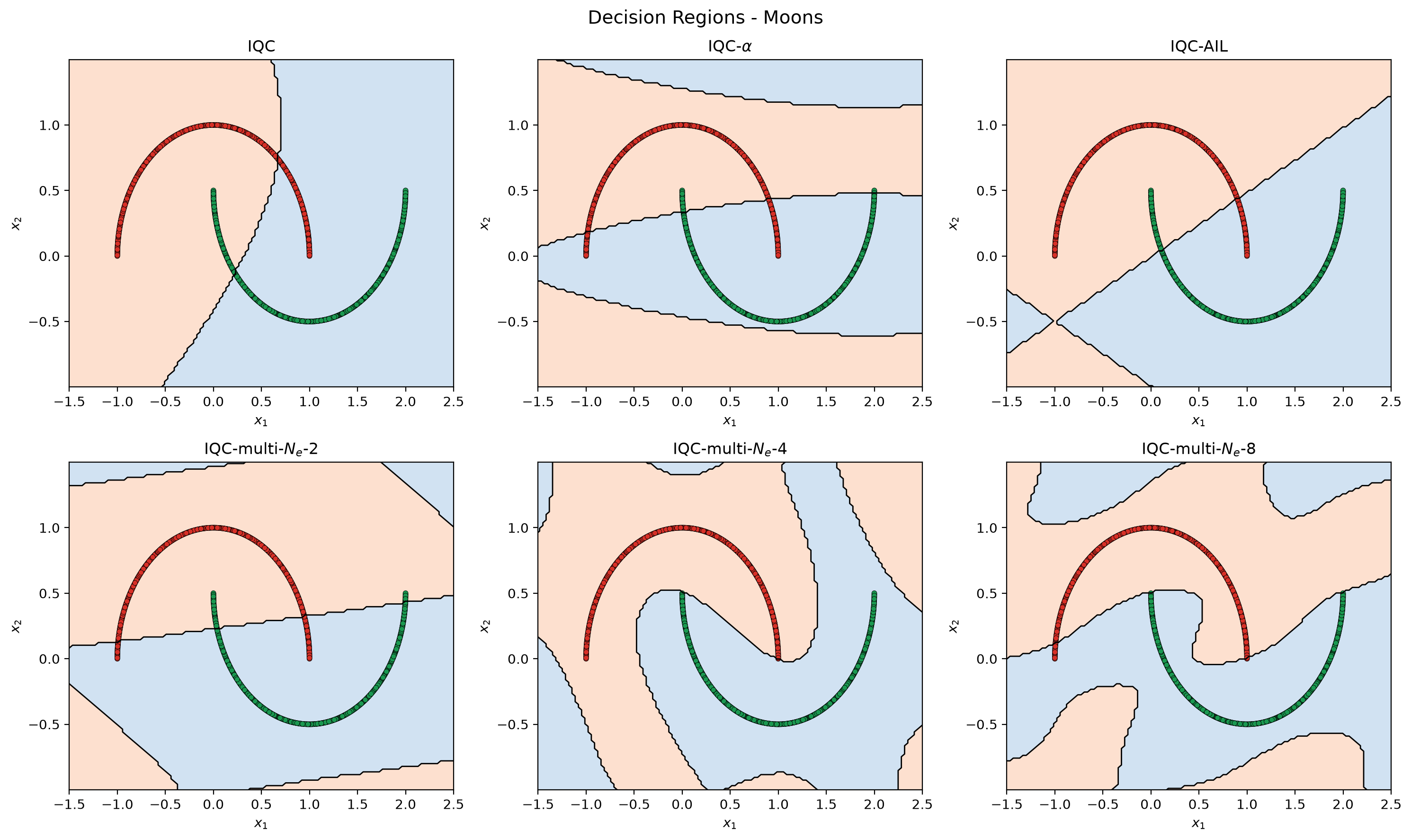}
    \caption{Decision regions learned by the six \IQC{} variants on the
    Moons dataset. The plots show the resulting classification regions
    together with the data points for the two classes.}
    \label{fig:decision_regions_moons}
\end{figure}

\begin{figure}[!htbp]
  \centering
  \includegraphics[width=\linewidth]{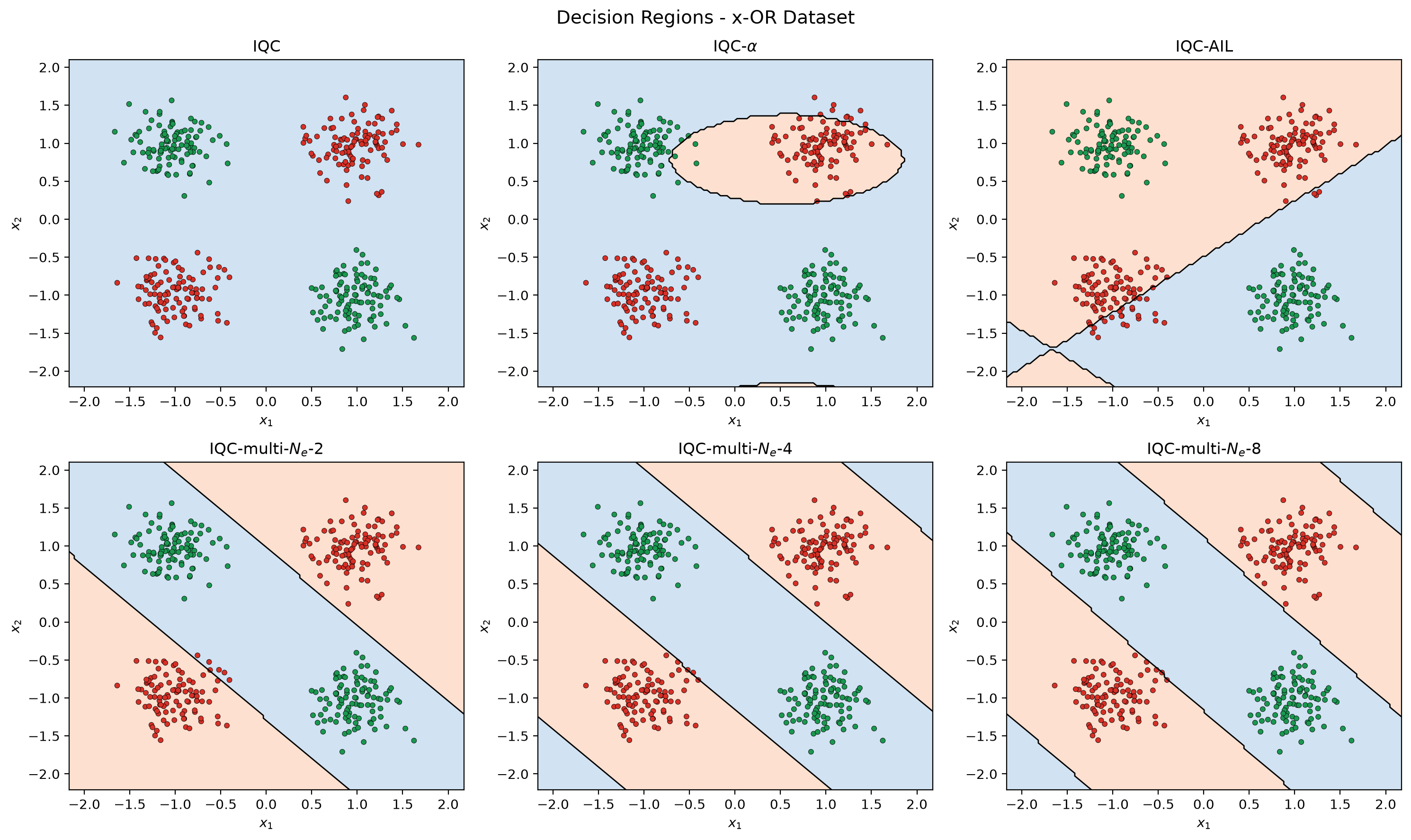}
  \caption{Decision regions learned by the six \IQC{} variants on the XOR
    dataset. The multidimensional variants generate structured
    alternating regions that separate the four data clusters.}
  \label{fig:decision_regions_xor}
\end{figure}

To complement the quantitative results, Fig.~\ref{fig:decision_regions_circle},
Fig.~\ref{fig:decision_regions_moons}, and
Fig.~\ref{fig:decision_regions_xor} visualize the decision regions
learned by the six \IQC{} variants on representative nonlinear datasets.
The figures provide a geometric view of the functions generated by the
different parametrizations.

On the Circles dataset, the baseline \IQC{} produces a relatively
simple decision structure that does not closely follow the circular
geometry of the data. The multidimensional variants produce
 richer decision regions, with the resulting boundaries
exhibiting structures that more closely follow the nonlinear geometry
of the dataset. A similar behavior is observed for Moons, where the
multidimensional models generate more structured decision regions than
the baseline and non-multidimensional parametrizations.

The difference is particularly evident for XOR. The non-multidimensional
variants produce decision regions that do not adequately separate the
four clusters, whereas the multidimensional variants generate multiple
alternating regions that successfully separate the classes. This
qualitative observation is consistent with the substantially higher
classification performance of the multidimensional models on this
dataset.

These decision regions provide a geometric interpretation of the
different functional structures generated by the models. They should
not be interpreted as a direct visualization of individual Fourier
components; rather, they represent the resulting prediction function
obtained from the superposition of the frequency components described
in Sec.~\ref{sec:fourier}. In this sense, the richer spatial
structures observed for the multidimensional variants are consistent
with the richer set of multidimensional frequencies enabled by the
environment Hamiltonian. This is discussed in more detail in
Sec.~\ref{sec:discussion}.



\subsection{Expressibility}
\label{sec:expressibility}

We next characterize the expressibility of the state ensembles
generated by the different \IQC{} variants using the fidelity-distribution
framework described in Sec.~\ref{sec:evaluation-metrics}. Since the Haar
reference distribution depends explicitly on the global Hilbert-space
dimension, expressibility values obtained for different dimensions
cannot be directly interpreted as a ranking of the architectures. We
therefore analyze both the fidelity distributions relative to their
corresponding Haar references and idle-normalized comparisons
using the relative expressibility measure introduced in
Sec.~\ref{sec:evaluation-metrics}.

\begin{figure}[!htbp]
    \centering
    \includegraphics[width=\textwidth]{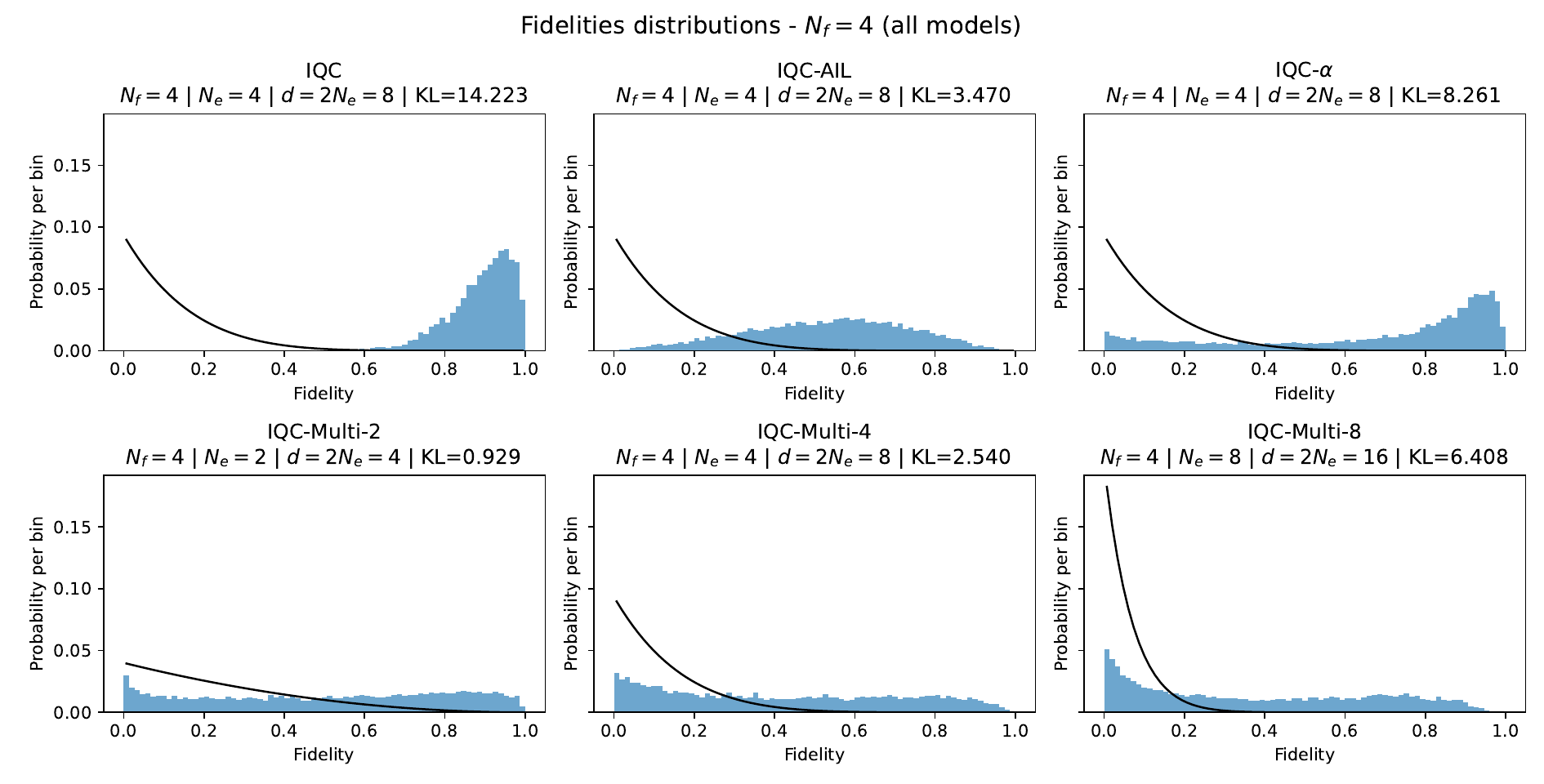}
    \caption{Empirical fidelity distributions for the six \IQC{} models at
    $N_f=4$, together with the corresponding Haar reference
    distributions. The values of $N_e$, the global Hilbert-space
    dimension, and the KL divergence are indicated for each model.}
    \label{fig:fidelity_distributions_nf4}
\end{figure}

For $N_f=4$, Fig.~\ref{fig:fidelity_distributions_nf4} shows the
empirical fidelity distributions for all six models together with their
corresponding Haar reference distributions. The models exhibit
substantially different distances from their respective Haar ensembles.
Among the \IQCmulti{} variants, \IQCmulti[2] presents the smallest
KL divergence. However, these three models correspond to different
global Hilbert-space dimensions, $d=4$, $8$, and $16$, respectively.
Consequently, their KL values should not be interpreted as a
dimension-independent ranking of the architectures. Instead, the
figure illustrates how each parametrization compares with the Haar
distribution associated with its own Hilbert-space dimension.

\begin{figure}[!htbp]
    \centering
    \includegraphics[width=\textwidth]{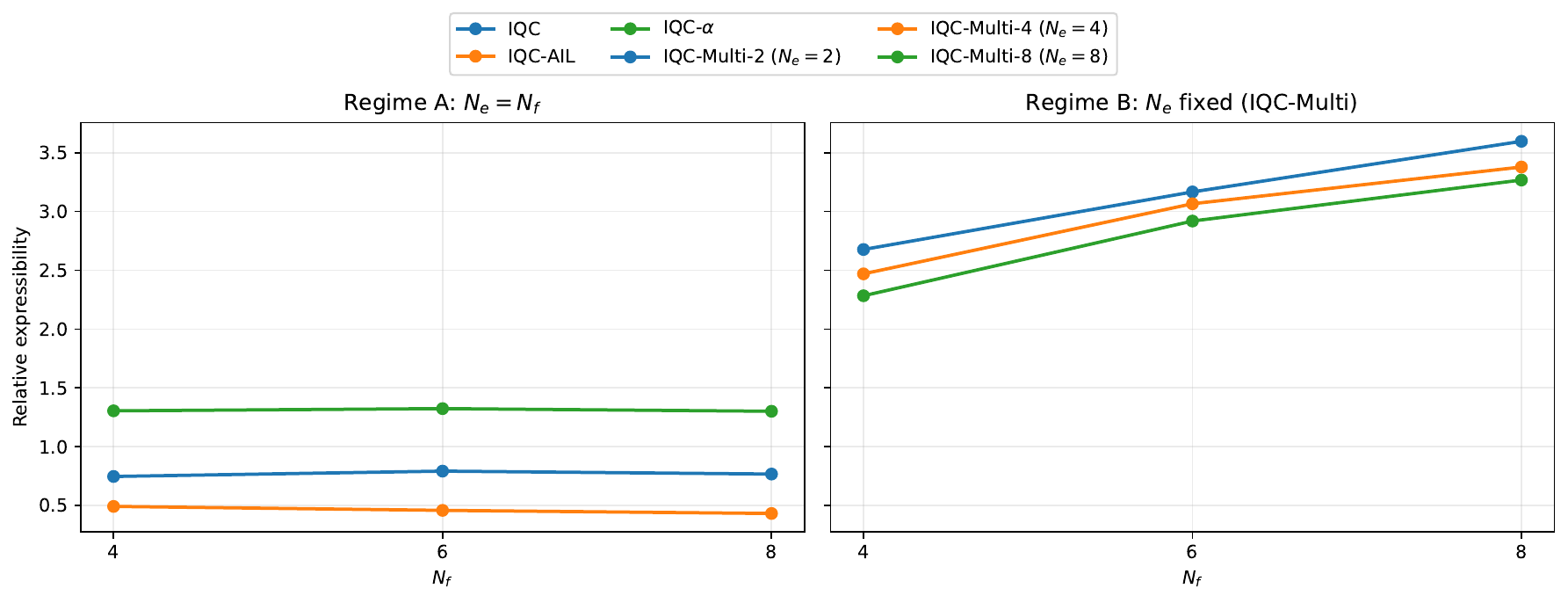}
    \caption{Relative expressibility as a function of the number of input
    features $N_f$. Left: \IQC{}, \IQCail{}, and \IQCalpha{} with
    $N_e=N_f$. Right: \IQCmulti{} models with fixed environment dimensions
    $N_e=2$, $4$, and $8$.}
    \label{fig:expressibility_scaling}
\end{figure}

A complementary analysis is obtained by studying how relative
expressibility scales with the number of input
features. Fig.~\ref{fig:expressibility_scaling} shows the relative
expressibility $\mathrm{RExpr}$ defined in
Eq.~\eqref{eq:relative_expr}. This normalization allows comparisons
between architectures with different Hilbert-space dimensions by
measuring the reduction of the KL divergence relative to the
idle-circuit baseline.

For the \IQC{}, \IQCail{}, and \IQCalpha{} parametrizations considered
in our benchmark, the environment dimension is increased together with
the number of input features, $N_e=N_f$. In this regime, the KL
divergence increases with $N_f$. However, this increase occurs
simultaneously with an increase of the global Hilbert-space dimension
and therefore with a change in the corresponding Haar reference. For
each of these parametrizations, the relative expressibility, in
contrast, remains approximately constant over the investigated range
of feature dimensions. This indicates that increasing the number of
features while proportionally increasing the environment dimension
does not significantly change the expressibility relative to the
corresponding idle baseline.

The \IQCmulti{} architectures exhibit a different scaling behavior. In
this case, the environment dimension is fixed for each model while the
number of input features is varied. Therefore, the global
Hilbert-space dimension and the Haar reference distribution remain
unchanged within each \IQCmulti{} family. The results show a
systematic increase of the relative expressibility with $N_f$ for all
investigated environment dimensions. Since this increase occurs
without enlarging the Hilbert space available to the model, it cannot
be attributed simply to an increase in the number of qubits. Instead,
it is associated with the ability of the \IQCmulti{} Hamiltonian
parametrization to generate state ensembles that become closer to the
Haar reference as additional input dimensions are incorporated into
the environmental encoding.

\begin{figure}[!htbp]
    \centering
    \includegraphics[width=0.85\textwidth]{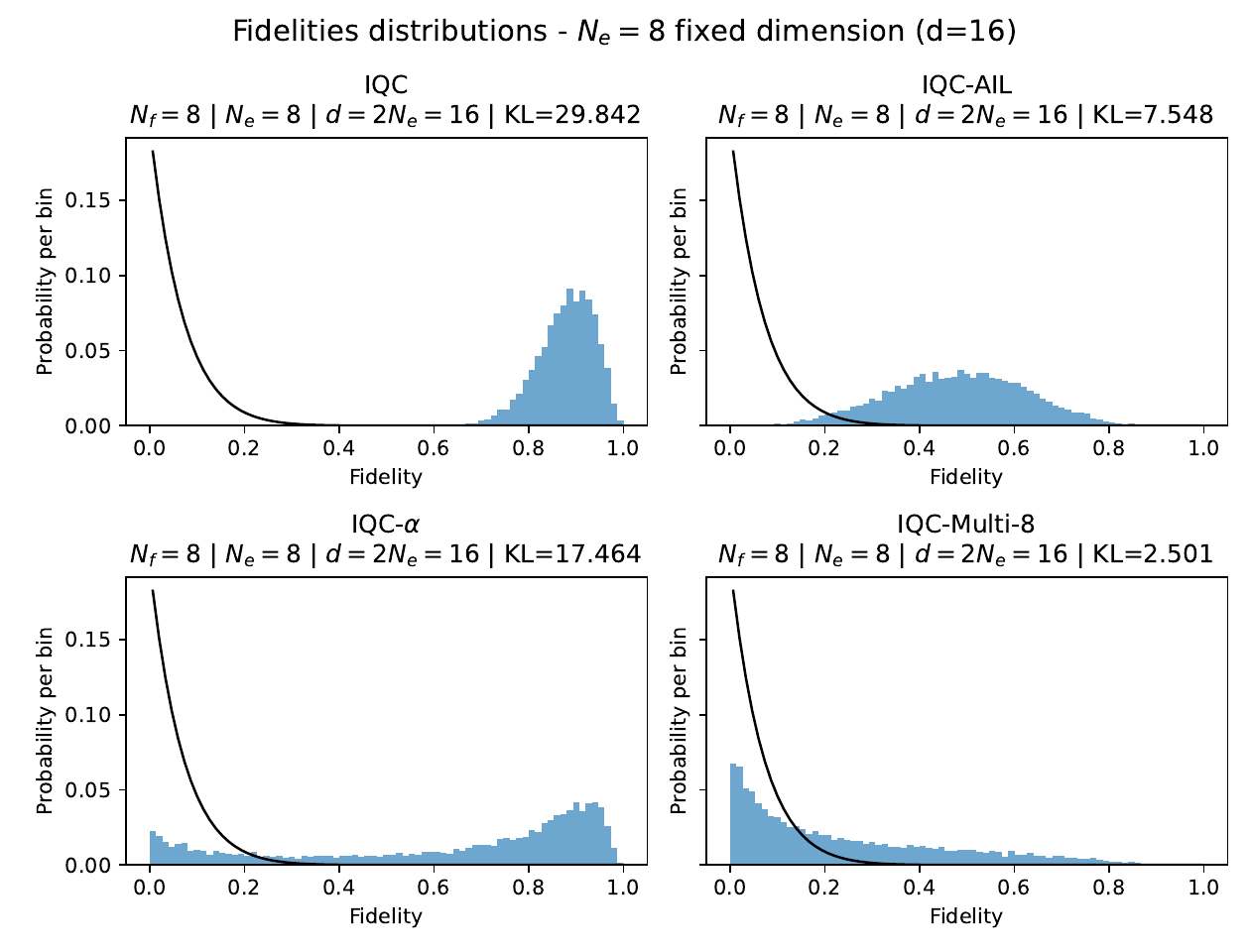}
    \caption{Fidelity distributions for the four models with $N_f=8$
      and the same global Hilbert-space dimension, $d=16$, together
      with the common Haar reference distribution.}
    \label{fig:fidelity_distributions_nf8_fixed_d16}
\end{figure}

A controlled architectural comparison can finally be performed by fixing
both the input dimension and the global Hilbert-space dimension. For
$N_f=8$, the models \IQC{}, \IQCail{},
\IQCalpha{}, and \IQCmulti[8] all have $N_e=8$ and
therefore $d=16$. Their fidelity distributions are shown in
Fig.~\ref{fig:fidelity_distributions_nf8_fixed_d16}, where all models
are compared with the same Haar reference distribution.

Under this controlled comparison, \IQCmulti[8] exhibits the smallest
KL divergence, followed by \IQCail{}, \IQCalpha{}, and \IQC{}. Because
all models are evaluated against the same Haar distribution, this
comparison provides the most direct assessment of the effect of the
Hamiltonian parametrization on the generated state ensemble. The
substantially smaller KL divergence of \IQCmulti[8] indicates that this
architecture generates states with the closest fidelity distribution
to the Haar ensemble among the models considered at fixed global
dimension.

The expressibility results therefore reveal two distinct effects. On
one hand, increasing the global Hilbert-space dimension does not
necessarily lead to increased relative expressibility, as observed for
the models where $N_e=N_f$. On the other hand, modifying the structure
of the environmental Hamiltonian through \IQCmulti{} can increase the
relative expressibility even when the environment dimension is kept
fixed. This highlights that expressibility is determined not only by
the available Hilbert-space dimension, but also by how the input
features are incorporated into the Hamiltonian structure.

Our expressibility measure characterizes the statistical structure of
the global state ensemble generated by an architecture, whereas
classification performance reflects the subset of this ensemble that
can be reached through training and that provides a useful
representation for a particular dataset. The relationship between
these two notions and their implications for IQC performance are
discussed in the next section.

 \section{Discussion}\label{sec:discussion}

 The results in Sec.~\ref{sec:results} establish a direct connection
 between the Hamiltonian parametrization of the \IQC{} and the Fourier
 structure of the resulting prediction function. Rather than revealing
 a simple hierarchy in which increasingly expressive models
 systematically outperform simpler ones, the results indicate that
 predictive performance depends on the compatibility between the
 accessible Fourier structure and the structure of the underlying
 dataset. This interpretation follows naturally from the Fourier
 characterization of parametrized quantum models, in which the
 encoding determines the accessible frequencies while the trainable
 circuit structure and measurement determine the corresponding
 coefficients \citep{schuldEffectDataEncoding2021}. Related work on
 learning Fourier functions with parametrized quantum circuits further
 emphasizes the role of accessible spectral structure in determining
 the functions that can be represented
 \citep{heimannLearningFourierSeries2025}.

At the classification level, the investigated parametrizations provide
different types of spectral degrees of freedom. As discussed in
Sec.~\ref{sec:iqc-models}, \IQCalpha{} increases the freedom of the
Fourier coefficients through a trainable target Hamiltonian, whereas
\IQCmulti{} allows the environmental eigenvalues to depend on linear
combinations of input features, providing multidimensional control of
the accessible frequencies. The classification results show that the
\IQCmulti{} family achieves the strongest aggregate performance among
the considered parametrizations, with \IQCmulti[8] obtaining the
highest aggregate score (Table~\ref{tab:overall-classification}).
However, the pairwise analysis in
Table~\ref{tab:wilcoxon-holm-f1-overall} shows that several
differences are not statistically significant, and the results
therefore do not support interpreting \IQCmulti{} as universally
superior. Instead, they suggest that additional spectral degrees of
freedom are beneficial when they provide a Fourier structure
compatible with the target learning problem.
 
This behavior can be understood from the Fourier representation derived
in Sec.~\ref{sec:fourier}. In \IQCalpha{}, the additional trainable
parameters are introduced through the target Hamiltonian and therefore
primarily modify the Fourier coefficients while preserving the
feature-wise structure of the frequencies. The resulting prediction
function retains the separable form
\begin{equation}
\label{eq:f-separable}
f_{\text{\IQCalpha{}}}(\btheta,\bx)
=
\sum_{l=0}^{N_f-1}\tilde{f}(\bhalpha,aw_lx_l).
\end{equation}
Thus, \IQCalpha{} provides greater control over the combination of
existing Fourier components without introducing general couplings
between input features.

In contrast, \IQCmulti{} retains the same coefficient structure as
\IQCalpha{}, while allowing each environmental eigenvalue to depend on
a linear combination of input features:
\begin{equation}
\lambda_l(\bm{x})=\bm{w}_l^T\bm{x}.
\end{equation}
The resulting Fourier representation is therefore non-separable in the
original feature coordinates, providing a richer set of accessible
multidimensional frequencies. The distinction can consequently be
viewed as one between coefficient control and spectral-structure
control. This interpretation is consistent with recent analyses
showing that the presence of a frequency in the accessible spectrum
does not necessarily imply that its coefficient can be controlled
independently or efficiently, due to frequency redundancy and
coefficient
correlations~\citep{mhiriConstrainedVanishingExpressivity2025,stroblFourierFingerprintsAnsatzes2025}.

The \IQC{} formulation makes this distinction particularly explicit in
our prediction function \eqref{eq:experimental-prediction}. The
direction of the target Hamiltonian, $\hat{\bm{\alpha}}$, determines
the Fourier coefficients, whereas its effective scale $\kappa$ in
Eq.~\eqref{eq:kappa-parameter} rescales the accessible
frequencies. Thus, the target Hamiltonian simultaneously controls the
spectral scale and the corresponding coefficients, suggesting that
future reparametrizations separating these two roles may provide more
controlled optimization landscapes. This structure introduces
substantial parameter sharing across the $N_{e}$ frequency components,
since their coefficients are constrained by the three components of
the normalized target direction $\bhalpha$. This redundancy can
potentially be mitigated by a multi-qubit target Hamiltonian.

The expressibility analysis in Sec.~\ref{sec:expressibility} provides
a complementary perspective.  Because the Haar reference depends on
the global Hilbert-space dimension, changes in expressibility must be
interpreted carefully when the environment dimension varies (see
Fig.~\ref{fig:fidelity_distributions_nf4}). For \IQC{}, \IQCail{}, and
\IQCalpha{}, $N_e=N_f$, so increasing the number of features also
increases the Hilbert-space dimension. The approximately constant
relative expressibility observed in this regime, as seen in
Fig.~\ref{fig:expressibility_scaling}, indicates that the increase in
KL divergence does not correspond to a substantial change in the
normalized distance from the Haar ensemble.

\IQCmulti{} allows $N_f$ and $N_e$ to be varied independently. Its
increasing relative expressibility with $N_f$ at fixed environment
dimension therefore shows that improved agreement with the Haar
reference need not result from an increase in the number of qubits.
The controlled comparison at $N_f=8$ and $d=16$ in
Fig.~\ref{fig:fidelity_distributions_nf8_fixed_d16} further shows that
\IQCmulti[8] has the smallest KL divergence among the models sharing
the same global dimension. Together, these results indicate that the
structure of the environmental Hamiltonian plays an important role in
determining the generated state ensemble.

Nevertheless, global state-ensemble expressibility does not directly
predict classification performance. Haar-based expressibility
characterizes the statistical structure of the global
target--environment state ensemble, whereas classification depends on
whether the accessible Fourier structure provides a useful
representation for the particular learning problem. Task-relevant
Fourier expressivity and global state expressibility should therefore
be regarded as complementary rather than equivalent notions. This
distinction is also consistent with the observation that increased
expressibility does not necessarily imply improved
trainability~\citep{holmes2022connecting}.

The decision regions shown in Sec.~\ref{sec:decision-regions} provide
a geometric illustration of these differences (see
Figs.~\ref{fig:decision_regions_circle},
\ref{fig:decision_regions_moons} and
\ref{fig:decision_regions_xor}). The qualitatively different
boundaries generated by the parametrizations reflect differences in
the accessible frequencies and in the coefficients with which those
frequencies are combined. The richer spatial structures observed for
the multidimensional variants are therefore consistent with their
richer multidimensional Fourier structure.

The decision boundaries can be obtained by setting
\eqref{eq:experimental-prediction}, for \IQCmulti{}, and
\eqref{eq:exp-ail-prediction-2}, for \IQCail{}, equal to zero. For
\IQCmulti{}, we can understand the decision boundaries as a result of
a superposition of Fourier components with different multidimensional
frequencies. This argument justifies why we see a diagonal wave-like
structure in the XOR dataset (see the bottom line in
Fig.~\ref{fig:decision_regions_xor}) and the complex wave-like regions
in the Moons dataset (see the bottom line in
Fig.~\ref{fig:decision_regions_moons}). Meanwhile, the quadratic
function of \IQCail{} in \eqref{eq:exp-ail-prediction-2} explains the
cone-like decision lines in Figs.~\ref{fig:decision_regions_moons} and
\ref{fig:decision_regions_xor}.

An important structural advantage of \IQCmulti{} is that the feature
dimension and environment dimension can be independently controlled.
This provides an additional degree of freedom for shaping the
accessible spectral structure without requiring a proportional increase
in the number of input features. However, the classification results
show that increasing this resource does not guarantee improved
performance, indicating that the appropriate spectral complexity is
dataset-dependent.

The \IQCail{} model should be interpreted separately from this
Hamiltonian hierarchy because it changes the data-encoding mechanism
by incorporating the input into the initial environment state. Its
distinct behavior suggests that state-based and Hamiltonian-based
encodings constitute complementary sources of expressive power, and
their systematic interplay remains an interesting direction for future
work.

The analytical framework also suggests a complementary scaling
picture.  The environment dimension primarily controls the number and
structure of accessible frequency components, whereas the target
Hamiltonian controls their coefficients. Extending this picture to
multiple target qubits could therefore provide additional degrees of
freedom for Fourier coefficient control. A systematic study of this
relation and of Fourier-controllability measures could provide a more
direct characterization of the effective function class of \IQC{}s.

Finally, the present expressibility analysis characterizes the global
target--environment state ensemble, whereas the classifier output
depends only on the reduced target state after tracing out the
environment. The adopted metric may therefore capture degrees of
freedom that do not directly contribute to the prediction function.
Future studies should investigate reduced-state or observable-based
expressibility measures that more directly reflect the effective
function class implemented by the classifier.

 \section{Conclusion}\label{sec:conclusion}

 In this work, we introduced a general Hamiltonian formulation of
 Interactive Quantum Classifiers (\IQC{}s) and established their
 connection with the Fourier representation of parametrized quantum
 models. By treating the \IQC{} as a Hamiltonian-induced reduced
 quantum channel, we derived the target dynamics analytically and
 showed that the resulting prediction function admits an explicit
 Fourier expansion. This formulation reveals a natural separation
 between the roles of the two interacting subsystems: the
 environmental Hamiltonian determines the input-dependent spectral
 structure, whereas the target Hamiltonian controls how these spectral
 components contribute to the observable output.

 Building upon this framework, we analyzed different Hamiltonian
 parametrizations of \IQC{}s, including the previously introduced
 \IQCalpha{} model and the current \IQCmulti{} generalization. The
 latter allows the environment dimension to vary independently of the
 feature dimension, providing additional freedom in the accessible
 Fourier structure. Across the considered benchmarks, the \IQCmulti{}
 family achieved the strongest aggregate classification performance,
 although the results do not imply a universal ordering among
 models. Instead, predictive performance depends on the compatibility
 between the accessible Fourier structure and the learning task.

 The expressibility analysis further shows that global state-ensemble
 expressibility and predictive performance capture different
 properties of quantum learning models. Richer Hamiltonian
 parametrizations can produce state ensembles closer to the Haar
 reference according to the adopted metric without necessarily
 improving classification. These results emphasize the distinction
 between global state expressibility and task-relevant Fourier
 expressivity, in which both accessible frequencies and the
 controllability of their associated coefficients play a central role.

 The present formulation also suggests several directions for future
 work. The frequency scale is controlled by the scalar factor
 $\kappa$, which depends on the target-system parameters $\alpha_0$
 and $\bm{\alpha}$, whereas the Fourier coefficients depend on the
 normalized target direction $\hat{\bm{\alpha}}$. Independently
 parameterizing these quantities may provide more direct spectral
 control and lead to improved optimization strategies. Increasing the
 number of target qubits may likewise provide additional degrees of
 freedom for coefficient control, suggesting a complementary scaling
 picture in which the environment dimension primarily controls
 spectral capacity while the target dimension may provide additional
 coefficient capacity. Further work could investigate
 Fourier-controllability measures, the interplay between state-based
 and Hamiltonian-based encodings, and task-relevant expressibility
 measures based on reduced target states or observable
 outputs. Extending the analytically tractable formulation to
 hardware-efficient and scalable implementations is another natural
 direction.

\subsubsection*{Acknowledgements}

FMPN acknowledges financial support from the Brazilian National
Council for Scientific and Technological Development (CNPq) under
Grant No.~408499/2025-7.  FMPN is also a researcher affiliated with
the National Institute of Science and Technology for Applied Quantum
Computing, supported by CNPq under Grant No.~408884/2024-0 and part of
the Brazilian National Institute for Artificial Intelligence,
supported by CNPq under Grant No.~406417/2022-9.

\subsubsection*{Data Availability}
\label{sec:data-availability}

The source code and numerical results supporting the findings of this
work are publicly available at
\url{https://github.com/QuICS-Lab/parametrized_interactive_quantum_classifiers}.
The real-world datasets used in this study are publicly available from
their respective original sources, with scripts provided in the
repository to reproduce their acquisition and preprocessing.

\appendix

\section{Classification Results}
\label{sec:class-results}

In this section, we present all the complete training results per
metric per model summarized in Sec.~\ref{sec:results}.

\begin{table}[!htbp]
\tiny
\centering
\caption{Results (mean $\pm$ standard deviation) of the IQC models on
  the binary classification datasets.  }
\label{tab:results_binary}
\begin{tabular}{llccccc}
\toprule
Dataset & Model & Accuracy & AUC & F1-score & Precision & Recall \\
\midrule
\multirow{6}{*}{Blobs 2D} & \IQC{} & \textbf{1.0000 $\pm$ 0.0000} & \textbf{1.0000 $\pm$ 0.0000} & \textbf{1.0000 $\pm$ 0.0000} & \textbf{1.0000 $\pm$ 0.0000} & \textbf{1.0000 $\pm$ 0.0000} \\
 & \IQCalpha{} & \textbf{1.0000 $\pm$ 0.0000} & \textbf{1.0000 $\pm$ 0.0000} & \textbf{1.0000 $\pm$ 0.0000} & \textbf{1.0000 $\pm$ 0.0000} & \textbf{1.0000 $\pm$ 0.0000} \\
 & \IQCail{} & \textbf{1.0000 $\pm$ 0.0000} & \textbf{1.0000 $\pm$ 0.0000} & \textbf{1.0000 $\pm$ 0.0000} & \textbf{1.0000 $\pm$ 0.0000} & \textbf{1.0000 $\pm$ 0.0000} \\
 & \IQCmulti[2] & \underline{0.9967 $\pm$ 0.0070} & \underline{0.9972 $\pm$ 0.0060} & \underline{0.9966 $\pm$ 0.0073} & 0.9961 $\pm$ 0.0083 & \underline{0.9972 $\pm$ 0.0060} \\
 & \IQCmulti[4] & 0.9933 $\pm$ 0.0211 & 0.9946 $\pm$ 0.0171 & 0.9931 $\pm$ 0.0217 & 0.9926 $\pm$ 0.0234 & 0.9946 $\pm$ 0.0171 \\
 & \IQCmulti[8] & \underline{0.9967 $\pm$ 0.0070} & 0.9970 $\pm$ 0.0064 & \underline{0.9966 $\pm$ 0.0072} & \underline{0.9963 $\pm$ 0.0079} & 0.9970 $\pm$ 0.0064 \\
\cmidrule(lr){2-7}
\multirow{6}{*}{Blobs 4D} & \IQC{} & \textbf{1.0000 $\pm$ 0.0000} & \textbf{1.0000 $\pm$ 0.0000} & \textbf{1.0000 $\pm$ 0.0000} & \textbf{1.0000 $\pm$ 0.0000} & \textbf{1.0000 $\pm$ 0.0000} \\
 & \IQCalpha{} & \textbf{1.0000 $\pm$ 0.0000} & \textbf{1.0000 $\pm$ 0.0000} & \textbf{1.0000 $\pm$ 0.0000} & \textbf{1.0000 $\pm$ 0.0000} & \textbf{1.0000 $\pm$ 0.0000} \\
 & \IQCail{} & \textbf{1.0000 $\pm$ 0.0000} & \textbf{1.0000 $\pm$ 0.0000} & \textbf{1.0000 $\pm$ 0.0000} & \textbf{1.0000 $\pm$ 0.0000} & \textbf{1.0000 $\pm$ 0.0000} \\
 & \IQCmulti[2] & \underline{0.9983 $\pm$ 0.0053} & \underline{0.9985 $\pm$ 0.0048} & \underline{0.9983 $\pm$ 0.0053} & \underline{0.9982 $\pm$ 0.0056} & \underline{0.9985 $\pm$ 0.0048} \\
 & \IQCmulti[4] & \textbf{1.0000 $\pm$ 0.0000} & \textbf{1.0000 $\pm$ 0.0000} & \textbf{1.0000 $\pm$ 0.0000} & \textbf{1.0000 $\pm$ 0.0000} & \textbf{1.0000 $\pm$ 0.0000} \\
 & \IQCmulti[8] & \textbf{1.0000 $\pm$ 0.0000} & \textbf{1.0000 $\pm$ 0.0000} & \textbf{1.0000 $\pm$ 0.0000} & \textbf{1.0000 $\pm$ 0.0000} & \textbf{1.0000 $\pm$ 0.0000} \\
\cmidrule(lr){2-7}
\multirow{6}{*}{Blobs 8D} & \IQC{} & \textbf{1.0000 $\pm$ 0.0000} & \textbf{1.0000 $\pm$ 0.0000} & \textbf{1.0000 $\pm$ 0.0000} & \textbf{1.0000 $\pm$ 0.0000} & \textbf{1.0000 $\pm$ 0.0000} \\
 & \IQCalpha{} & \textbf{1.0000 $\pm$ 0.0000} & \textbf{1.0000 $\pm$ 0.0000} & \textbf{1.0000 $\pm$ 0.0000} & \textbf{1.0000 $\pm$ 0.0000} & \textbf{1.0000 $\pm$ 0.0000} \\
 & \IQCail{} & \textbf{1.0000 $\pm$ 0.0000} & \textbf{1.0000 $\pm$ 0.0000} & \textbf{1.0000 $\pm$ 0.0000} & \textbf{1.0000 $\pm$ 0.0000} & \textbf{1.0000 $\pm$ 0.0000} \\
 & \IQCmulti[2] & \underline{0.9967 $\pm$ 0.0105} & \underline{0.9970 $\pm$ 0.0096} & \underline{0.9967 $\pm$ 0.0106} & \underline{0.9966 $\pm$ 0.0109} & \underline{0.9970 $\pm$ 0.0096} \\
 & \IQCmulti[4] & 0.9850 $\pm$ 0.0372 & 0.9850 $\pm$ 0.0371 & 0.9850 $\pm$ 0.0372 & 0.9850 $\pm$ 0.0372 & 0.9850 $\pm$ 0.0371 \\
 & \IQCmulti[8] & 0.9917 $\pm$ 0.0264 & 0.9917 $\pm$ 0.0262 & 0.9917 $\pm$ 0.0264 & 0.9917 $\pm$ 0.0264 & 0.9917 $\pm$ 0.0262 \\
\cmidrule(lr){2-7}
\multirow{6}{*}{Circles} & \IQC{} & 0.5640 $\pm$ 0.0408 & 0.5692 $\pm$ 0.0316 & 0.5435 $\pm$ 0.0858 & 0.5489 $\pm$ 0.1139 & 0.5692 $\pm$ 0.0316 \\
 & \IQCalpha{} & 0.6980 $\pm$ 0.0657 & 0.6983 $\pm$ 0.0665 & 0.6966 $\pm$ 0.0653 & 0.7024 $\pm$ 0.0689 & 0.6983 $\pm$ 0.0665 \\
 & \IQCail{} & 0.4890 $\pm$ 0.0193 & 0.4915 $\pm$ 0.0119 & 0.4549 $\pm$ 0.0503 & 0.4814 $\pm$ 0.1228 & 0.4915 $\pm$ 0.0119 \\
 & \IQCmulti[2] & 0.7180 $\pm$ 0.0877 & 0.7183 $\pm$ 0.0875 & 0.7173 $\pm$ 0.0877 & 0.7203 $\pm$ 0.0883 & 0.7183 $\pm$ 0.0875 \\
 & \IQCmulti[4] & \underline{0.7395 $\pm$ 0.0546} & \underline{0.7398 $\pm$ 0.0547} & \underline{0.7389 $\pm$ 0.0551} & \underline{0.7486 $\pm$ 0.0746} & \underline{0.7398 $\pm$ 0.0547} \\
 & \IQCmulti[8] & \textbf{0.8080 $\pm$ 0.0713} & \textbf{0.8088 $\pm$ 0.0699} & \textbf{0.8064 $\pm$ 0.0726} & \textbf{0.8162 $\pm$ 0.0659} & \textbf{0.8088 $\pm$ 0.0699} \\
\cmidrule(lr){2-7}
\multirow{6}{*}{Moons} & \IQC{} & 0.7505 $\pm$ 0.0334 & 0.7512 $\pm$ 0.0335 & 0.7500 $\pm$ 0.0333 & 0.7532 $\pm$ 0.0342 & 0.7512 $\pm$ 0.0335 \\
 & \IQCalpha{} & 0.8340 $\pm$ 0.0453 & 0.8344 $\pm$ 0.0465 & 0.8334 $\pm$ 0.0463 & 0.8374 $\pm$ 0.0451 & 0.8344 $\pm$ 0.0465 \\
 & \IQCail{} & 0.8560 $\pm$ 0.0228 & 0.8560 $\pm$ 0.0228 & 0.8558 $\pm$ 0.0228 & 0.8567 $\pm$ 0.0229 & 0.8560 $\pm$ 0.0228 \\
 & \IQCmulti[2] & 0.8380 $\pm$ 0.0442 & 0.8385 $\pm$ 0.0449 & 0.8377 $\pm$ 0.0447 & 0.8399 $\pm$ 0.0444 & 0.8385 $\pm$ 0.0449 \\
 & \IQCmulti[4] & \underline{0.8810 $\pm$ 0.0626} & \underline{0.8812 $\pm$ 0.0629} & \underline{0.8808 $\pm$ 0.0628} & \underline{0.8820 $\pm$ 0.0620} & \underline{0.8812 $\pm$ 0.0629} \\
 & \IQCmulti[8] & \textbf{0.9290 $\pm$ 0.0405} & \textbf{0.9289 $\pm$ 0.0405} & \textbf{0.9289 $\pm$ 0.0406} & \textbf{0.9304 $\pm$ 0.0412} & \textbf{0.9289 $\pm$ 0.0405} \\
\cmidrule(lr){2-7}
\multirow{6}{*}{Stripes} & \IQC{} & 0.9000 $\pm$ 0.0323 & 0.9009 $\pm$ 0.0325 & 0.8997 $\pm$ 0.0325 & 0.9017 $\pm$ 0.0316 & 0.9009 $\pm$ 0.0325 \\
 & \IQCalpha{} & 0.8925 $\pm$ 0.0302 & 0.8932 $\pm$ 0.0299 & 0.8923 $\pm$ 0.0302 & 0.8945 $\pm$ 0.0323 & 0.8932 $\pm$ 0.0299 \\
 & \IQCail{} & 0.8750 $\pm$ 0.0276 & 0.8753 $\pm$ 0.0283 & 0.8744 $\pm$ 0.0279 & 0.8785 $\pm$ 0.0263 & 0.8753 $\pm$ 0.0283 \\
 & \IQCmulti[2] & \underline{0.9225 $\pm$ 0.0520} & \underline{0.9232 $\pm$ 0.0500} & \underline{0.9223 $\pm$ 0.0521} & \underline{0.9252 $\pm$ 0.0507} & \underline{0.9232 $\pm$ 0.0500} \\
 & \IQCmulti[4] & \textbf{0.9300 $\pm$ 0.0494} & \textbf{0.9302 $\pm$ 0.0493} & \textbf{0.9299 $\pm$ 0.0495} & \textbf{0.9300 $\pm$ 0.0496} & \textbf{0.9302 $\pm$ 0.0493} \\
 & \IQCmulti[8] & 0.9137 $\pm$ 0.0602 & 0.9146 $\pm$ 0.0598 & 0.9135 $\pm$ 0.0604 & 0.9149 $\pm$ 0.0591 & 0.9146 $\pm$ 0.0598 \\
\cmidrule(lr){2-7}
\multirow{6}{*}{XOR} & \IQC{} & 0.4787 $\pm$ 0.0981 & 0.4807 $\pm$ 0.0745 & 0.4479 $\pm$ 0.0576 & 0.4860 $\pm$ 0.0864 & 0.4807 $\pm$ 0.0745 \\
 & \IQCalpha{} & 0.5012 $\pm$ 0.1023 & 0.5082 $\pm$ 0.0769 & 0.4777 $\pm$ 0.0323 & 0.4951 $\pm$ 0.0400 & 0.5082 $\pm$ 0.0769 \\
 & \IQCail{} & 0.5487 $\pm$ 0.1308 & 0.5771 $\pm$ 0.1328 & 0.5316 $\pm$ 0.1493 & 0.5599 $\pm$ 0.1975 & 0.5771 $\pm$ 0.1328 \\
 & \IQCmulti[2] & 0.9850 $\pm$ 0.0219 & 0.9854 $\pm$ 0.0217 & 0.9848 $\pm$ 0.0221 & 0.9851 $\pm$ 0.0216 & 0.9854 $\pm$ 0.0217 \\
 & \IQCmulti[4] & \textbf{0.9963 $\pm$ 0.0084} & \textbf{0.9964 $\pm$ 0.0083} & \textbf{0.9962 $\pm$ 0.0085} & \textbf{0.9962 $\pm$ 0.0083} & \textbf{0.9964 $\pm$ 0.0083} \\
 & \IQCmulti[8] & \underline{0.9925 $\pm$ 0.0134} & \underline{0.9909 $\pm$ 0.0169} & \underline{0.9921 $\pm$ 0.0143} & \underline{0.9938 $\pm$ 0.0108} & \underline{0.9909 $\pm$ 0.0169} \\
\bottomrule
\end{tabular}
\vspace{5pt}
\begin{minipage}{0.98\textwidth}\footnotesize
\textbf{Bold} indicates the best result and \underline{underlining} indicates the second-best result for each dataset and metric.
\end{minipage}
\end{table}

\begin{table}[!htbp]
\tiny
\centering
\caption{Results (mean $\pm$ standard deviation) of the IQC models on
  the real-world classification datasets. }
\label{tab:results_realworld}
\begin{tabular}{llcccc}
\toprule
Dataset & Model & Accuracy & F1-score & Precision & Recall \\
\midrule
\multirow{6}{*}{Caesarian} & \IQC{} & 0.5125 $\pm$ 0.1208 & 0.3718 $\pm$ 0.1568 & 0.3731 $\pm$ 0.2486 & 0.5206 $\pm$ 0.1100 \\
 & \IQCalpha{} & 0.5250 $\pm$ 0.0604 & 0.4543 $\pm$ 0.1005 & \textbf{0.6310 $\pm$ 0.2185} & \underline{0.5762 $\pm$ 0.0509} \\
 & \IQCail{} & 0.5312 $\pm$ 0.0896 & 0.4336 $\pm$ 0.1214 & 0.5138 $\pm$ 0.1979 & 0.5373 $\pm$ 0.0686 \\
 & \IQCmulti[2] & \textbf{0.5437 $\pm$ 0.1064} & \textbf{0.5266 $\pm$ 0.1146} & \underline{0.5921 $\pm$ 0.1393} & 0.5722 $\pm$ 0.1087 \\
 & \IQCmulti[4] & 0.5188 $\pm$ 0.1064 & \underline{0.4886 $\pm$ 0.1181} & 0.5848 $\pm$ 0.1662 & 0.5532 $\pm$ 0.1013 \\
 & \IQCmulti[8] & \underline{0.5375 $\pm$ 0.1222} & 0.4814 $\pm$ 0.1631 & 0.5647 $\pm$ 0.2137 & \textbf{0.5794 $\pm$ 0.1088} \\
\cmidrule(lr){2-6}
\multirow{6}{*}{Iris} & \IQC{} & 0.9100 $\pm$ 0.0473 & 0.9089 $\pm$ 0.0482 & 0.9194 $\pm$ 0.0449 & 0.9100 $\pm$ 0.0473 \\
 & \IQCalpha{} & 0.9100 $\pm$ 0.0498 & 0.9094 $\pm$ 0.0504 & 0.9165 $\pm$ 0.0481 & 0.9100 $\pm$ 0.0498 \\
 & \IQCail{} & 0.7333 $\pm$ 0.0754 & 0.7206 $\pm$ 0.0589 & 0.7457 $\pm$ 0.0752 & 0.7333 $\pm$ 0.0754 \\
 & \IQCmulti[2] & \underline{0.9367 $\pm$ 0.0838} & 0.9333 $\pm$ 0.0906 & 0.9389 $\pm$ 0.0820 & \underline{0.9367 $\pm$ 0.0838} \\
 & \IQCmulti[4] & \textbf{0.9433 $\pm$ 0.0417} & \underline{0.9423 $\pm$ 0.0428} & \textbf{0.9502 $\pm$ 0.0358} & \textbf{0.9433 $\pm$ 0.0417} \\
 & \IQCmulti[8] & \textbf{0.9433 $\pm$ 0.0498} & \textbf{0.9428 $\pm$ 0.0507} & \underline{0.9456 $\pm$ 0.0484} & \textbf{0.9433 $\pm$ 0.0498} \\
\cmidrule(lr){2-6}
\multirow{6}{*}{Pima diabetes} & \IQC{} & 0.6494 $\pm$ 0.0000 & 0.3937 $\pm$ 0.0000 & 0.3247 $\pm$ 0.0000 & 0.5000 $\pm$ 0.0000 \\
 & \IQCalpha{} & 0.6500 $\pm$ 0.0021 & 0.3957 $\pm$ 0.0062 & 0.3749 $\pm$ 0.1588 & 0.5028 $\pm$ 0.0088 \\
 & \IQCail{} & 0.6494 $\pm$ 0.0000 & 0.3937 $\pm$ 0.0000 & 0.3247 $\pm$ 0.0000 & 0.5000 $\pm$ 0.0000 \\
 & \IQCmulti[2] & \textbf{0.6994 $\pm$ 0.0454} & \textbf{0.5353 $\pm$ 0.1157} & \textbf{0.6549 $\pm$ 0.2299} & \textbf{0.5781 $\pm$ 0.0688} \\
 & \IQCmulti[4] & \underline{0.6753 $\pm$ 0.0297} & \underline{0.4845 $\pm$ 0.0823} & \underline{0.6488 $\pm$ 0.1902} & \underline{0.5443 $\pm$ 0.0453} \\
 & \IQCmulti[8] & 0.6721 $\pm$ 0.0318 & 0.4636 $\pm$ 0.0913 & 0.5975 $\pm$ 0.2372 & 0.5362 $\pm$ 0.0501 \\
\cmidrule(lr){2-6}
\multirow{6}{*}{Wine} & \IQC{} & 0.7694 $\pm$ 0.0936 & 0.7352 $\pm$ 0.1226 & 0.8387 $\pm$ 0.1211 & 0.7379 $\pm$ 0.1051 \\
 & \IQCalpha{} & 0.8806 $\pm$ 0.0615 & 0.8806 $\pm$ 0.0679 & 0.9073 $\pm$ 0.0486 & 0.8763 $\pm$ 0.0667 \\
 & \IQCail{} & \textbf{0.9389 $\pm$ 0.0366} & \textbf{0.9409 $\pm$ 0.0355} & \textbf{0.9472 $\pm$ 0.0312} & \textbf{0.9404 $\pm$ 0.0373} \\
 & \IQCmulti[2] & 0.9139 $\pm$ 0.0462 & 0.9165 $\pm$ 0.0444 & 0.9194 $\pm$ 0.0432 & 0.9198 $\pm$ 0.0450 \\
 & \IQCmulti[4] & \underline{0.9222 $\pm$ 0.0552} & \underline{0.9214 $\pm$ 0.0572} & \underline{0.9264 $\pm$ 0.0536} & \underline{0.9237 $\pm$ 0.0575} \\
 & \IQCmulti[8] & 0.8889 $\pm$ 0.0898 & 0.8849 $\pm$ 0.0951 & 0.8985 $\pm$ 0.0925 & 0.8846 $\pm$ 0.0958 \\
\bottomrule
\end{tabular}
\vspace{5pt}
\begin{minipage}{0.98\textwidth}\footnotesize
  \textbf{Bold} indicates the best result and \underline{underlining}
  indicates the second-best result for each dataset and metric.
\end{minipage}
\end{table}

\clearpage{}
\section{Statistical Tests}
\label{app:wilcoxon-holm}

In this section, we present all the complete statistical results per
model summarized in Sec.~\ref{sec:results}.

\begin{table*}[!htbp]
\centering
\caption{Pairwise Wilcoxon signed-rank tests with Holm correction for the F1-score on the Circles dataset.}
\label{tab:wilcoxon-holm-f1-circle-no-noise}
\small
\setlength{\tabcolsep}{4pt}
\begin{tabular}{lcccccc}
\toprule
Model & \IQC{} & \IQCalpha{} & \IQCail{} & \IQCmulti[2] & \IQCmulti[4] & \IQCmulti[8] \\
\midrule
\IQC{} & -- & 0.0527 & 0.2578 & 0.0527 & 0.0293$^{\downarrow}$ & 0.0293$^{\downarrow}$ \\
\IQCalpha{} &  & -- & 0.0293$^{\uparrow}$ & 1.0000 & 0.6973 & 0.0957 \\
\IQCail{} &  &  & -- & 0.0293$^{\downarrow}$ & 0.0293$^{\downarrow}$ & 0.0293$^{\downarrow}$ \\
\IQCmulti[2] &  &  &  & -- & 1.0000 & 0.1172 \\
\IQCmulti[4] &  &  &  &  & -- & 0.1172 \\
\IQCmulti[8] &  &  &  &  &  & -- \\
\bottomrule
\end{tabular}
\vspace{2mm}
\begin{minipage}{0.98\textwidth}\footnotesize
Each cell reports the Holm-adjusted $p$-value. The symbol $\uparrow$ indicates that the model in the row achieved a significantly higher F1-score than the model in the column. The symbol $\downarrow$ indicates a significantly lower F1-score. Values without arrows indicate no statistically significant difference at $\alpha=0.05$.
\end{minipage}
\end{table*}

\begin{table*}[!htbp]
\centering
\caption{Pairwise Wilcoxon signed-rank tests with Holm correction for the F1-score on the Stripes dataset.}
\label{tab:wilcoxon-holm-f1-stripes}
\small
\setlength{\tabcolsep}{4pt}
\begin{tabular}{lcccccc}
\toprule
Model & \IQC{} & \IQCalpha{} & \IQCail{} & \IQCmulti[2] & \IQCmulti[4] & \IQCmulti[8] \\
\midrule
\IQC{} & -- & 1.0000 & 0.0547 & 1.0000 & 1.0000 & 1.0000 \\
\IQCalpha{} &  & -- & 0.2344 & 1.0000 & 0.1777 & 1.0000 \\
\IQCail{} &  &  & -- & 0.4082 & 0.0293$^{\downarrow}$ & 0.7812 \\
\IQCmulti[2] &  &  &  & -- & 1.0000 & 1.0000 \\
\IQCmulti[4] &  &  &  &  & -- & 1.0000 \\
\IQCmulti[8] &  &  &  &  &  & -- \\
\bottomrule
\end{tabular}
\vspace{2mm}
\begin{minipage}{0.98\textwidth}\footnotesize
Each cell reports the Holm-adjusted $p$-value. The symbol $\uparrow$ indicates that the model in the row achieved a significantly higher F1-score than the model in the column. The symbol $\downarrow$ indicates a significantly lower F1-score. Values without arrows indicate no statistically significant difference at $\alpha=0.05$.
\end{minipage}
\end{table*}

\begin{table*}[!htbp]
\centering
\caption{Pairwise Wilcoxon signed-rank tests with Holm correction for the F1-score on the Iris dataset.}
\label{tab:wilcoxon-holm-f1-iris}
\small
\setlength{\tabcolsep}{4pt}
\begin{tabular}{lcccccc}
\toprule
Model & \IQC{} & \IQCalpha{} & \IQCail{} & \IQCmulti[2] & \IQCmulti[4] & \IQCmulti[8] \\
\midrule
\IQC{} & -- & 1.0000 & 0.0293$^{\uparrow}$ & 1.0000 & 0.3516 & 0.2539 \\
\IQCalpha{} &  & -- & 0.0293$^{\uparrow}$ & 1.0000 & 1.0000 & 1.0000 \\
\IQCail{} &  &  & -- & 0.0293$^{\downarrow}$ & 0.0293$^{\downarrow}$ & 0.0293$^{\downarrow}$ \\
\IQCmulti[2] &  &  &  & -- & 1.0000 & 1.0000 \\
\IQCmulti[4] &  &  &  &  & -- & 1.0000 \\
\IQCmulti[8] &  &  &  &  &  & -- \\
\bottomrule
\end{tabular}
\vspace{2mm}
\begin{minipage}{0.98\textwidth}\footnotesize
Each cell reports the Holm-adjusted $p$-value. The symbol $\uparrow$ indicates that the model in the row achieved a significantly higher F1-score than the model in the column. The symbol $\downarrow$ indicates a significantly lower F1-score. Values without arrows indicate no statistically significant difference at $\alpha=0.05$.
\end{minipage}
\end{table*}

\begin{table*}[!htbp]
\centering
\caption{Pairwise Wilcoxon signed-rank tests with Holm correction for the F1-score on the Moons dataset.}
\label{tab:wilcoxon-holm-f1-moons-no-noise}
\small
\setlength{\tabcolsep}{4pt}
\begin{tabular}{lcccccc}
\toprule
Model & \IQC{} & \IQCalpha{} & \IQCail{} & \IQCmulti[2] & \IQCmulti[4] & \IQCmulti[8] \\
\midrule
\IQC{} & -- & 0.0469$^{\downarrow}$ & 0.0293$^{\downarrow}$ & 0.0391$^{\downarrow}$ & 0.0293$^{\downarrow}$ & 0.0293$^{\downarrow}$ \\
\IQCalpha{} &  & -- & 1.0000 & 1.0000 & 0.6328 & 0.0293$^{\downarrow}$ \\
\IQCail{} &  &  & -- & 0.9668 & 0.9668 & 0.0391$^{\downarrow}$ \\
\IQCmulti[2] &  &  &  & -- & 0.2598 & 0.0293$^{\downarrow}$ \\
\IQCmulti[4] &  &  &  &  & -- & 0.9668 \\
\IQCmulti[8] &  &  &  &  &  & -- \\
\bottomrule
\end{tabular}
\vspace{2mm}
\begin{minipage}{0.98\textwidth}\footnotesize
Each cell reports the Holm-adjusted $p$-value. The symbol $\uparrow$ indicates that the model in the row achieved a significantly higher F1-score than the model in the column. The symbol $\downarrow$ indicates a significantly lower F1-score. Values without arrows indicate no statistically significant difference at $\alpha=0.05$.
\end{minipage}
\end{table*}

\begin{table*}[!htbp]
\centering
\caption{Pairwise Wilcoxon signed-rank tests with Holm correction for the F1-score on the XOR dataset.}
\label{tab:wilcoxon-holm-f1-xor}
\small
\setlength{\tabcolsep}{4pt}
\begin{tabular}{lcccccc}
\toprule
Model & \IQC{} & \IQCalpha{} & \IQCail{} & \IQCmulti[2] & \IQCmulti[4] & \IQCmulti[8] \\
\midrule
\IQC{} & -- & 0.9375 & 0.9375 & 0.0293$^{\downarrow}$ & 0.0293$^{\downarrow}$ & 0.0293$^{\downarrow}$ \\
\IQCalpha{} &  & -- & 1.0000 & 0.0293$^{\downarrow}$ & 0.0293$^{\downarrow}$ & 0.0293$^{\downarrow}$ \\
\IQCail{} &  &  & -- & 0.0293$^{\downarrow}$ & 0.0293$^{\downarrow}$ & 0.0293$^{\downarrow}$ \\
\IQCmulti[2] &  &  &  & -- & 0.9375 & 1.0000 \\
\IQCmulti[4] &  &  &  &  & -- & 1.0000 \\
\IQCmulti[8] &  &  &  &  &  & -- \\
\bottomrule
\end{tabular}
\vspace{2mm}
\begin{minipage}{0.98\textwidth}\footnotesize
Each cell reports the Holm-adjusted $p$-value. The symbol $\uparrow$ indicates that the model in the row achieved a significantly higher F1-score than the model in the column. The symbol $\downarrow$ indicates a significantly lower F1-score. Values without arrows indicate no statistically significant difference at $\alpha=0.05$.
\end{minipage}
\end{table*}

\begin{table*}[!htbp]
\centering
\caption{Pairwise Wilcoxon signed-rank tests with Holm correction for the F1-score on the Pima Indians Diabetes dataset.}
\label{tab:wilcoxon-holm-f1-pima-diabetes}
\small
\setlength{\tabcolsep}{4pt}
\begin{tabular}{lcccccc}
\toprule
Model & \IQC{} & \IQCalpha{} & \IQCail{} & \IQCmulti[2] & \IQCmulti[4] & \IQCmulti[8] \\
\midrule
\IQC{} & -- & 1.0000 & 1.0000 & 0.1875 & 0.1172 & 0.2812 \\
\IQCalpha{} &  & -- & 1.0000 & 0.1875 & 0.1172 & 0.3281 \\
\IQCail{} &  &  & -- & 0.1875 & 0.1172 & 0.2812 \\
\IQCmulti[2] &  &  &  & -- & 0.6328 & 0.9668 \\
\IQCmulti[4] &  &  &  &  & -- & 1.0000 \\
\IQCmulti[8] &  &  &  &  &  & -- \\
\bottomrule
\end{tabular}
\vspace{2mm}
\begin{minipage}{0.98\textwidth}\footnotesize
Each cell reports the Holm-adjusted $p$-value. The symbol $\uparrow$ indicates that the model in the row achieved a significantly higher F1-score than the model in the column. The symbol $\downarrow$ indicates a significantly lower F1-score. Values without arrows indicate no statistically significant difference at $\alpha=0.05$.
\end{minipage}
\end{table*}

\begin{table*}[!htbp]
\centering
\caption{Pairwise Wilcoxon signed-rank tests with Holm correction for the F1-score on the Wine dataset.}
\label{tab:wilcoxon-holm-f1-wine}
\small
\setlength{\tabcolsep}{4pt}
\begin{tabular}{lcccccc}
\toprule
Model & \IQC{} & \IQCalpha{} & \IQCail{} & \IQCmulti[2] & \IQCmulti[4] & \IQCmulti[8] \\
\midrule
\IQC{} & -- & 0.1504 & 0.0293$^{\downarrow}$ & 0.0293$^{\downarrow}$ & 0.0508 & 0.1953 \\
\IQCalpha{} &  & -- & 0.1172 & 0.9668 & 0.7852 & 1.0000 \\
\IQCail{} &  &  & -- & 0.9668 & 1.0000 & 0.5195 \\
\IQCmulti[2] &  &  &  & -- & 1.0000 & 0.5156 \\
\IQCmulti[4] &  &  &  &  & -- & 0.4922 \\
\IQCmulti[8] &  &  &  &  &  & -- \\
\bottomrule
\end{tabular}
\vspace{2mm}
\begin{minipage}{0.98\textwidth}\footnotesize
Each cell reports the Holm-adjusted $p$-value. The symbol $\uparrow$ indicates that the model in the row achieved a significantly higher F1-score than the model in the column. The symbol $\downarrow$ indicates a significantly lower F1-score. Values without arrows indicate no statistically significant difference at $\alpha=0.05$.
\end{minipage}
\end{table*}

\clearpage{}

\bibliographystyle{sn-mathphys-num-jheppub}
\bibliography{qml-multi-v2}

\end{document}